\documentclass[preprint,1p,times]{elsarticle}

\usepackage{times}
\usepackage{graphicx}
\usepackage{footnote}
\usepackage{array}
\usepackage{comment}
\usepackage{graphicx}
\usepackage{placeins}
\usepackage{float}
\usepackage{subfigure}
\usepackage[table,xcdraw]{xcolor}
\usepackage{url}
\usepackage{graphicx}
\usepackage{array}
\usepackage[inline]{enumitem}
\usepackage[table]{xcolor}
\usepackage{tabularx}
\usepackage{caption}
\usepackage{float}
\usepackage{tabu}
\usepackage{multicol}
\usepackage{multirow}
\usepackage{longtable}
\usepackage{flushend}
\usepackage{adjustbox}
\usepackage{anyfontsize}
\usepackage{footnote}
\makesavenoteenv{tabular}
\makesavenoteenv{table}
\usepackage{lipsum}
\usepackage{courier}
\usepackage{adjustbox}
\usepackage{booktabs} 

\colorlet{tableheadcolor}{gray!25} 
\newcommand{\headcol}{\rowcolor{tableheadcolor}} %
\colorlet{tablerowcolor}{gray!10} 
\newcommand{\rowcol}{\rowcolor{tablerowcolor}} %
\usepackage{booktabs}
\usepackage{colortbl}
\usepackage{amsmath}
\usepackage{xcolor}
\usepackage{graphicx}
\colorlet{tableheadcolor}{gray!25} 
\colorlet{tablerowcolor}{gray!10} 
\newcommand{\topline}{\arrayrulecolor{black}\specialrule{0.1em}{\abovetopsep}{0pt}%
	\arrayrulecolor{tableheadcolor}\specialrule{\belowrulesep}{0pt}{0pt}%
	\arrayrulecolor{black}}
\newcommand{\midline}{\arrayrulecolor{tableheadcolor}\specialrule{\aboverulesep}{0pt}{0pt}%
	\arrayrulecolor{black}\specialrule{\lightrulewidth}{0pt}{0pt}%
	\arrayrulecolor{white}\specialrule{\belowrulesep}{0pt}{0pt}%
	\arrayrulecolor{black}}

\newcommand{\bottomline}{\arrayrulecolor{white}\specialrule{\aboverulesep}{0pt}{0pt}%
	\arrayrulecolor{black}\specialrule{\heavyrulewidth}{0pt}{\belowbottomsep}}%
\newcommand{\bottomlinec}{\arrayrulecolor{tablerowcolor}\specialrule{\aboverulesep}{0pt}{0pt}%
	\arrayrulecolor{black}\specialrule{\heavyrulewidth}{0pt}{\belowbottomsep}}%
\usepackage{amssymb}
\usepackage{amsmath}
\usepackage{hyperref}
\usepackage{url}

\journal{Intelligent Data Analysis}

\begin{document}

\begin{frontmatter}



\title{Hybrid Recommender Systems: A Systematic Literature Review}

\author{Erion \c Cano\fnref{cor1}}
\ead{erion.cano@polito.it} 	


\author{Maurizio Morisio}
\ead{maurizio.morisio@polito.it}

\fntext[cor1]{Corresponding Author}

\address{Department of Control and Computer Engineering, Politecnico di Torino, Corso 
Duca degli Abruzzi, 24 - 10129 Torino}





\begin{abstract}
Recommender systems are software tools used to generate and provide suggestions for
items and other entities to the users by exploiting various strategies. 
Hybrid recommender systems combine two or more recommendation strategies in different
ways to benefit from their complementary advantages. This systematic literature review
presents the state of the art in hybrid recommender systems of the last decade. It is 
the first quantitative review work completely focused in hybrid recommenders.
We address the most relevant problems considered and present the associated
data mining and recommendation techniques used to overcome them. We also explore
the hybridization classes each hybrid recommender belongs to, the application domains, 
the evaluation process and proposed future research directions. Based on our findings, 
most of the studies combine collaborative filtering with another technique often in 
a weighted way. Also cold-start and data sparsity are the two traditional and top problems 
being addressed in 23 and 22 studies each, while movies and movie datasets are still 
widely used by most of the authors. As most of the studies are evaluated by comparisons 
with similar methods using accuracy metrics, providing more credible and user 
oriented evaluations remains a typical challenge. Besides this, newer challenges 
were also identified such as responding to the variation of user context, evolving user 
tastes or providing cross-domain recommendations. Being a hot topic, hybrid recommenders 
represent a good basis with which to respond accordingly by exploring newer opportunities 
such as contextualizing recommendations, involving parallel hybrid algorithms, processing 
larger datasets, etc.
\end{abstract}


\end{frontmatter}


\vskip 5mm
\section{Introduction}
\label{intro}
Historically people have relied on their peers or on experts' suggestions for decision 
support and recommendations about commodities, news, entertainment, etc. The exponential 
growth of the digital information in the last 25 years, especially in the web, has created 
the problem of information overload. Information overload is defined as "stress induced 
by reception of more information than is necessary to make a decision and by attempts to 
deal with it with outdated time management practices".\footnote{\url{http://www.businessdictionary.com/definition/information-overload.html}}   
This problem limits our capacity to review the specifications and choose between numerous 
alternatives of items in the online market.  
On the other hand, information science and technology reacted accordingly by developing 
information filtering tools to alleviate the problem. Recommender Systems (RSs) are one 
such tools that emerged in the mid 90s. They are commonly defined as software tools and 
techniques used to provide suggestions for items and other recommendable entities to users 
\cite{Ricci2011}. In the early days (beginning of 90s) RSs were the study subject of other 
closely related research disciplines such as Human Computer Interaction (HCI) or Information 
Retrieval (IR) \cite{Ekstrand:2011:CFR:2185827.2185828}. Today, RSs are found everywhere 
helping users in searching for various types of items and services. They also serve as 
sales assistants for businesses increasing their profits. 
\par
Technically all RSs employ one or more recommendation strategies such as Content-Based 
Filtering (CBF), Collaborative Filtering (CF), Demographic Filtering (DF), 
Knowledge-Based Filtering (KBF), etc. described below: 
\begin{itemize}[leftmargin=0.25cm]
	%
	\setlength\itemsep{0em}
	\item \textbf{Collaborative Filtering}: The basic assumption of CF is that people who had similar 
	tastes in the past will also have similar tastes in the future. One of its earliest definitions is 
	"collaboration between people to help one another perform filtering by recording their reactions to 
	documents they read" \cite{Goldberg:1992:UCF:138859.138867}. This approach uses ratings or 
	other forms of user generated feedback to spot taste commonalities between groups of users 
	and then generates recommendations based on inter-user similarities \cite{Ekstrand:2011:CFR:2185827.2185828}. CF recommenders suffer from problems like cold-start 
	(new user or new item), "gray sheep" (users that do not fit in any taste cluster), etc.     
	\item \textbf{Content-Based Filtering}: CBF is based on the assumption that people who 
	liked items with certain attributes in the past, will like the same kind of items in the future 
	as well. It makes use of item features to compare the item with user profiles 
	and provide recommendations. Recommendation quality is limited by the selected features of 
	the recommended items. Same as CF, CBF suffer from the cold-start problem.
	\item \textbf{Demographic Filtering}: DF uses demographic data such as \emph{age}, 
	\emph{gender}, \emph{education}, etc. for identifying categories of users. It does not 
	suffer from the new user problem as is doesn't use ratings to provide recommendations. 
	However, it is difficult today to collect enough demographic information that is needed 
	because of online privacy concerns, limiting the utilization of DF. It is still combined 
	with other recommenders as a reinforcing technique for better quality. 
	\item \textbf{Knowledge-Based Filtering}: KBF uses knowledge about users and
	items to reason about what items meet the users' requirements, and generate recommendations
	accordingly \cite{burke2000knowledgebased}. A special type of KBFs are 
	constraint-based RSs which are capable to recommend complex items that are 
	rarely bought (i.e. cars or houses) and manifest important constrains for the user 
	(price) \cite{Felfernig:2008:CRS:1409540.1409544}. It is not possible to 
	successfully use CF or CBF in this domain of items as few user-system interaction data 
	are available (people rarely buy houses). 
\end{itemize}
\par
One of the earliest recommender systems was Tapestry, a manual CF 
mail system \cite{Goldberg:1992:UCF:138859.138867}. 
The first computerized RS prototypes also applied a collaborative filtering approach and 
emerged in mid 90s \cite{Resnick:1994:GOA:192844.192905,Hill:1995:REC:223904.223929}.
GroupLens was a CF recommendation engine for finding news articles. 
In \cite{Hill:1995:REC:223904.223929} the authors present a detailed analysis and evaluation of
the Bellcore video recommender algorithm and its implementation embedded in the 
Mosaic browser interface. Ringo used taste similarities to provide personalized music 
recommendations. Other prototypes like NewsWeeder and InfoFinder recommended news and 
documents using CBF, based on item attributes \cite{lang95newsweeder,krulwich1996learning}. 
In late 90s important commercial RS prototypes also came out with Amazon.com recommender 
being the most popular. Many researchers started to combine the recommendation strategies in different ways building hybrid RSs which we consider in this review. Hybrid RSs put 
together two or more of the other strategies with the goal of reinforcing their 
advantages and reducing their disadvantages or limitations.	
One of the first was Fab, a meta-level recommender (see section 3.4.6) which  
was used to suggest websites \cite{Balabanovic:1997:FCC:245108.245124}. It incorporated 
a combination of CF to find users having similar website preferences, with CBF to find 
websites with similar content. Other works such as \cite{Sarwar:1998:UFA:289444.289509} 
followed shortly and hybrid RSs became a well established recommendation approach. 
\par
The continuously growing industrial interest in the recent and promising domains of 
mobile and social web has been followed by a similar increase of academic interest 
in RSs. ACM RecSys annual conference\footnote{\url{https://recsys.acm.org/}} is now the most 
significant event for presenting and discussing RS research. 
The work of Burke in \cite{Burke:2002:HRS:586321.586352} is one of the first qualitative 
surveys addressing hybrid RSs. The author analyzes advantages and 
disadvantages of the different recommendation strategies and provides a 
comprehensive taxonomy for classifying the ways they combine with each other to form 
hybrid RSs. He also presents several hybrid RS prototypes falling into the 7
hybridization classes of the taxonomy. Another early exploratory work is
\cite{Good:1999:CCF:315149.315352} where several experiments combining personalized
agents with opinions of community members in a CF framework are conducted. They conclude 
that this combination produces high-quality recommendations and that the best results of 
CF are achieved using large data of user communities. 
Other review works are more generic and address RSs 
in general, not focusing in any RS type. They reflect the increasing interest in the field 
in quantitative terms. In \cite{Bobadilla:2013:RSS:2483330.2483573} the authors perform a 
review work of 249 journal and conference RS publications from 1995 to 2013. The peak 
publication period of the works they consider is between 2007 and 2013 (last one-third of 
the analyzed period). They emphasize the fact that the current hybrid
RSs are incorporating location information into existing recommendation algorithms. They also 
highlight the proper combination of existing methods using different forms of data, and 
evaluating other characteristics (e.g., diversity and novelty) besides accuracy as 
future trends. In \cite{Park:2012:LRC:2181339.2181690} the authors review 210 recommender 
system articles published in 46 journals from 2001 to 2010. They similarly report a rapid 
increase of publications between 2007 and 2010 and predict an increase interest in mixing 
existing recommendation methods or using social network analysis to provide recommendations. 
\par
In this review paper we summarize the state of the art of hybrid RSs in the last 
10 years. We follow a systematic methodology to analyze and interpret the available 
facts related to the 7 research questions we defined. This methodology defined at 
\cite{kitchenham2004procedures, kitchenham2007guidelines} provides an unbiased and reproducible 
way for undertaking a review work. Unlike the other review works not focused in 
any RS type \cite{Bobadilla:2013:RSS:2483330.2483573, Park:2012:LRC:2181339.2181690},
this systematic literature review is the first quantitative work that is  
entirely focused in recent hybrid RS publications. For this reason it was not possible 
for us to have a direct basis with which to compare our results. Nevertheless we provide 
some comparisons of results for certain aspects in which hybrid RSs do not differ from 
other types of RSs. 
To have a general idea about what percentage of total RS publications address hybrid RSs
we examined \cite{Jannach2012}, a survey work about RSs in general. Here the 
authors review the work of 330 papers published in computer science and information 
systems conferences proceedings and journals from 2006 to 2011. Their results show 
that hybrid recommendation paradigm is the study object of about 14.5\% of their 
reviewed literature.
\par
We considered the most relevant problems hybrid RSs attempt to solve, 
the data mining and machine learning methods involved, RS technique combinations the 
studies utilize and the hybridization classes the proposed systems fall into. We also 
observed the domains in which the contributions were applied and the evaluation 
strategies, characteristics and metrics that were used. Based on the suggestions of 
the authors and the identified challenges we also present some future work directions 
which seem promising and in concordance with the RS trends.
Many primary studies were retrieved from digital libraries and the most relevant  
papers were selected for more detailed processing (we use the terms paper 
and study interchangeably to refer to the same object / concept). We hope this work 
will help anyone working in the field of (hybrid) RSs, especially by providing 
insights about future trends or opportunities.
The remainder of the paper is structured as follows. Section 2 briefly summarizes the 
methodology we followed, the objectives and research questions defined, the selection 
of papers and the quality assessment process. Section 3 introduces the results of the 
review organized in accordance with each research question. Section 4 discusses and 
summarizes each result whereas Section 5 concludes. Finally we list the selected papers 
in Appendix A.
\vskip 7mm
\section{Methodology}
The review work of this paper follows the guidelines that were defined by 
Kitchenham and Charters \cite{kitchenham2007guidelines} for systematic literature reviews 
in Software Engineering.   
\begin{figure}[!htbp]
	\centering
	\includegraphics[width=68mm,scale=0.5]{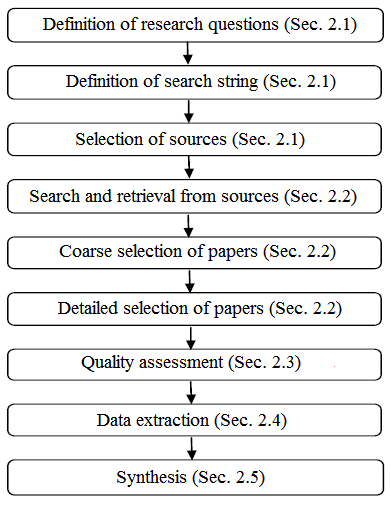}
	\caption{Systematic literature review protocol}
\end{figure}
The purpose of a systematic literature review is to present a verifiable and unbiased 
treatment of a research topic utilizing a rigorous and reproducible methodology. The 
guidelines that were followed are high level and do not consider the influence of research 
questions type on the review procedures. In Figure 1 we present the protocol of the review. 
It represents a clear set of steps which assist the management of the review process.
The protocol was defined by the first author and verified by the second author. In the 
following sections we describe each step we summarized in Figure 1. 		
\vskip 5mm
\subsection{Research questions, search string and digital sources}
The primary goal of this systematic literature review is to understand what challenges hybrid 
RSs could successfully address, how they are developed and evaluated and in what ways or 
aspects they could be experimented with. To this end, we defined the following research 
questions: 
\begin{description}
	\item[\textbf{RQ1}\hspace{1.45mm}] What are the most relevant studies addressing hybrid 
	recommender systems? 
	\item[\textbf{RQ2}\hspace{1.45mm}] What problems and challenges are faced by the researchers 
	in this field?
	\item[\textbf{RQ3a}] Which data mining and machine learning techniques are used in hybrid RSs?
	\item[\textbf{RQ3b}] What recommendation techniques are combined and which problems they solve?
	\item[\textbf{RQ4}\hspace{1.45mm}] What hybridization classes are used, based on the 
	taxonomy of Burke?	 
	\item[\textbf{RQ5}\hspace{1.45mm}] In what domains are hybrid recommenders applied?
	\item[\textbf{RQ6a}] What methodologies are used for the evaluation and what metrics they utilize?	
	\item[\textbf{RQ6b}] Which RS characteristics are evaluated and what metrics they use?
	\item[\textbf{RQ6c}] What datasets are used for training and testing hybrid RSs?
	\item[\textbf{RQ7}\hspace{1.45mm}] Which directions are most promising for future research?
\end{description}
\vskip 3mm
Furthermore we picked five scientific digital libraries that represent our primary sources for 
computer science research publications. They are listed in Table 1. Other similar sources were 
not considered as they mainly index data from the primary sources. 
\begin{table}[ht] 
	\caption{Selected sources to search for primary studies} 
	\small 
	\centering     
	\begin{tabular}
		{l l}   
		\topline
		\headcol \textbf{Source} & \textbf{URL}  \\ [0.5ex] 
		\midline   
		SpringerLink & http://link.springer.com   \\    
		\rowcol Science Direct & http://www.sciencedirect.com   \\ 
		IEEExplore & http://ieeexplore.ieee.org         \\ 
		\rowcol ACM Digital Library & http://dl.acm.org         \\ 
		Scopus & http://www.scopus.com                  \\ [0.5ex]  
		\bottomlinec
	\end{tabular}  
\end{table} 
\begin{table}[ht] 
	\caption{Keywords and synonyms} 
	\small 
	\centering     
	\begin{tabular}
		{l l}   
		\topline
		\headcol \textbf{Keyword} & \textbf{Synonyms}  \\ [0.5ex] 
		\midline   
		Hybrid & Hybridization, Mixed   \\   
		\rowcol Recommender & Recommendation  \\ 
		System & Software, Technique, Technology, Approach, Engine     \\ 
		\bottomlinec
	\end{tabular}  
\end{table}
We defined \textit{("Hybrid", "Recommender", "Systems")} as the basic set of keywords. 
Then we added synonyms to extend it and obtain the final set of keywords. The set of 
keywords and synonyms is listed in Table 2. The search string we defined is: 		\\ \\
\textit{
	("Hybrid" OR "Hybridization" OR "Mixed") AND 		
	("Recommender" OR "Recommendation")	AND ("System"	 
	OR "Software" OR "Technique" OR "Technology" OR 	
	"Approach" OR "Engine")}  
\vskip 5mm
\subsection{Selection of papers}
Following Step 4 of the protocol, we applied the search string in the search engines of 
the five digital libraries and found 9673 preliminary primary studies (see Table 4). The 
digital libraries return different numbers of papers because of the dissimilar filtering settings 
they use in their search engines. This retrieval process was conducted during May 2015. 
To objectively decide whether to select each preliminary primary study for further processing 
or not, we defined a set of inclusion / exclusion criteria listed in Table 3. 
The inclusion / exclusion criteria are considered as a basis of concentrating 
in the most relevant studies with which to achieve the objectives of the review.
\begin{table}[ht] 
	\caption{Inclusion and exclusion criteria}
	\footnotesize
	\centering      
	\begin{tabular}
		{p{9.7cm}}   
		\topline
		\headcol \textbf{Inclusion criteria}  		\\ [0.5ex] %
		\midline
		Papers presenting hybrid recommender systems, algorithms, approaches, etc. 			\\ [0.8ex]
		\rowcol Papers that even though do not specifically present hybrid RSs, provide		\\  		
		\rowcol recommendations combining different data mining techniques.								\\ [0.8ex]							
		Papers from conferences and journals       					 	 					\\ [0.8ex]
		\rowcol Papers published from 2005 to 2015              		 					\\ [0.8ex]
		Papers written in English language only                 		 					\\ [0.8ex]
		\hline
		\headcol \textbf{Exclusion criteria}  		\\
		\hline  
		Papers not addressing recommender systems at all   					 				\\ [0.8ex]
		\rowcol Papers addressing RSs but not implying any hybridization or combination		\\		
		\rowcol of different approaches or data mining techniques.		 			 		\\ [0.8ex]	   
		Papers that report only abstracts or slides of presentation,  		 			
		lacking detailed information 		   												\\ [0.8ex]
		\rowcol Grey literature                                 		   					\\ [0.8ex]
		\bottomline 
	\end{tabular}    
\end{table}
Duplicate papers were removed and a coarse selection 
phase followed. Processing all of them strictly was not practical. Therefore we 
decided to include journal and conference papers only, leaving out gray literature, 
workshop presentations or papers that report abstracts or presentation 
slides. We initially analyzed title, publication year and publication type 
(journal, conference, workshop, etc.). In many cases abstract or even more parts of each 
paper were examined for deciding to keep it or not. Our focus in this review work is 
on hybrid recommender systems. Thus we selected papers presenting mixed or combined 
RSs dropping out any paper addressing single recommendation strategies or papers not 
addressing RSs at all. Hybrid RSs represent a somehow newer family of recommender 
systems compared to other well known and widely used families such as CF or CBF. 
Therefore the last decade (2005-2015) was considered an appropriate publication period.
\begin{table}[ht]
	\caption{Number of papers after each selection step}
	\footnotesize
	\centering      
	\begin{tabular}{l c c c }
		\topline
		\headcol & \multicolumn{3}{c}{\textbf{Number of papers at the end of step:}}  \\ 		
		\cline{2-4} 
		\headcol \textbf{Digital source} & \textbf{Search and retrieval} & \textbf{Coarse selection} 
		& \textbf{Detailed selection}   \\
		\midline
		\multicolumn{1}{l}{SpringerLink} & 4152 & 50 & 13   					\\
		\rowcol \multicolumn{1}{l}{Scopus} & 3582 & 27 & 9  					\\
		\multicolumn{1}{l}{ACM Digital Library} & 1012 & 53 & 13  					\\
		\rowcol \multicolumn{1}{l}{Science Direct} & 484 & 35 & 12   				\\
		\multicolumn{1}{l}{IEEExplore} & 443 & 75 & 29							\\
		\hline
		\rowcol	\multicolumn{1}{l}{\textbf{Total}} & \textbf{9673} 
		& \textbf{240} & \textbf{76}   			\\
		\bottomline
	\end{tabular}
\end{table} 
Using inclusion / exclusion and this coarse selection step we reached to a list of 
240 papers. In the next step we performed a more detailed analysis and selection of 
the papers reviewing abstract and other parts of every paper. Besides relevance based 
on the inclusion / exclusion criteria, completeness (in terms of problem definition, 
description of the proposed method / technique / algorithm and evaluation of results) 
of each study was also taken into account. Finally we reached to our set of 76 
included papers. The full list is presented in Appendix A together with the 
publication details. 
\vskip 5mm
\subsection{Quality assessment}
We also defined 6 questions listed in Table 5 for the quality estimation of the selected 
studies. Each of the question receives score values of 0, 0.5 and 1 which represent answers "no", 
"partly" and "yes" correspondingly. The questions we defined do not reflect equal level 
of importance in the overall quality of the studies. For this reason we decided to 
weight them with coefficients of 0.5 (low importance) 1 (medium importance) and 1.5 
(high importance). 
\begin{table}[ht]
	\caption{Quality assessment questions} 
	\footnotesize
	\centering     
	\begin{tabular}
		{p{5.8cm} c r}   
		\topline
		\headcol \textbf{Quality Question} & \textbf{Score} & \textbf{Weight}  \\ [0.5ex] 
		\midline
		QQ1. Did the study clearly describe the 
		problems that is addressing? & yes / partly / no (1 / 0.5 / 0) & 1 \\     [0.8ex]         
		\rowcol QQ2. Did the study review the related work 
		for the problems? & yes / partly / no (1 / 0.5 / 0) & 0.5      \\	[0.8ex]
		QQ3. Did the study recommend any further 
		research? &	yes / partly / no (1 / 0.5 / 0) & 0.5 \\	[0.8ex]
		\rowcol QQ4. Did the study describe the components 
		or architecture of the proposed system? & yes / partly / no (1 / 0.5 / 0) & 1.5  \\	[0.8ex]
		QQ5. Did the study provide an empirical 
		evaluation? & yes / partly / no (1 / 0.5 / 0) & 1.5 \\	[0.8ex]
		\rowcol QQ6. Did the study present a clear statement 
		of findings? & yes / partly / no (1 / 0.5 / 0) & 1  \\		[0.8ex]
		\bottomline
	\end{tabular} 
\end{table}
We set higher weight to the quality questions that address the 
components/architecture of the system/solution (QQ4) and the empirical evaluation (QQ5). 
Quality questions that address problem description (QQ1) and statement of results (QQ6) 
got medium importance. We set a low importance weight to the two questions that address the 
related studies (QQ2) and future work (QQ3). The papers were split in two disjoint 
subsets. Each subset of papers was evaluated by one of the authors. In cases of 
indecision the quality score was set after a discussion between the authors. 
At the end, the final weighted quality score of each study was computed using the 
following formula: 
$$score = \sum_{i=1}^{6} w_{i} * v_{i} / 6$$
$w_{i}$ is the weight of question $i$ \textit{(0.5, 1, 1.5)} 		\\
$v_{i}$ is the vote for question $i$ \textit{(0, 0.5, 1)} 			\\	\\	
After this evaluation, cross-checking of the assessment was done on arbitrary studies 
(about 40\% of included papers) by the second author. At the end, an agreement on
differences was reached by discussion.
\vskip 5mm
\subsection{Data extraction }
Data extraction was carried on the final set of selected primary studies. We collected 
both paper meta-data (i.e., author, title, year, etc.) and content data important to answer 
our research questions like problems, application domains, etc. Table 6 presents 
our data extraction form. In the first column we list the extracted data, in the second
column we provide an explanation for some of the extracted data which may seem unclear 
and in the third column the research question with which the data is related. All the 
extracted information was stored in 
Nvivo\footnote{\url{http://www.qsrinternational.com/products.aspx}} 
which was used to manage data extraction and synthesis process. Nvivo is a data analysis 
software tool that helps in automating the identification and the labeling of the initial 
segments of text from the selected studies. 
\begin{table}[ht] 
	\caption{Data extraction form} 
	\scriptsize
	\centering      
	\begin{tabular}
		{l l l}  
		\topline
		\headcol \textbf{Extracted Data} & \textbf{Explanation} & \textbf{RQ}  				\\ [0.5ex]  
		\midline
		ID & A unique identifier of the form Pxx we set to each paper & -					\\
		\rowcol Title & - & RQ1																		\\
		Authors & - & -																		\\
		\rowcol Publication year & - & RQ1															\\
		Conference year & - & -																\\
		\rowcol Volume & Volume of the journal & -													\\
		Location & Location of the conference & -													\\
		\rowcol  Source & Digital library from which was retrieved & -								\\
		Publisher & - & -																	\\
		\rowcol Examiner & Name of person who performed data extraction & -							\\
		Participants & Study participants like students, academics, etc. 					\\
		\rowcol Goals & Work objectives & -															\\
		Application domain & Domain in which the study is applied & RQ5								\\
		\rowcol Approach & Hybrid recommendation approach applied & RQ3b							\\
		Contribution & Contribution of the research work & -								\\
		\rowcol Dataset & Public dataset used to train and evaluate the algorithm & RQ6c			\\
		DM techniques & Data mining techniques used & RQ3a									\\
		\rowcol Evaluation methodology & Methodology used to evaluate the RS & RQ6a					\\
		Evaluated characteristic & RS characteristics evaluated & RQ6b 								\\
		\rowcol Future work & Suggested future works & RQ7											\\
		Hybrid class & Class of hybrid RS & RQ4												\\
		\rowcol Research problem & - & RQ2															\\
		Score & Overall weighted quality score & - 											\\
		\rowcol Other Information & - & -															\\
		\bottomline 
	\end{tabular}  
\end{table}
\vskip 5mm
\subsection{Synthesis}
For the synthesis step we followed Cruzes and Dyba methodology for the thematic 
synthesis \cite{6092576}. Their methodology uses the concept of codes which are 
labeled segments of text to organize and aggregate the extracted information. 
Following the methodology we defined some initial codes which reflected the 
research questions. Some examples include the first research problems found, 
hybrid recommendation classes, first application domains, data mining techniques, 
recommendation approaches and evaluation methodologies. After completing the reading 
we had refined or detailed each of the initial codes with more precise sub-codes 
(leaf nodes in NVivo) which were even closer to the content of the selected papers, 
covering all the problems found, all the datasets used, and similar detailed data we found. 
We finished assigning codes to all the highlighted text segments of the papers and then
the codes were aggregated in themes (of different levels if necessary) by which 
the papers were grouped. Afterwards a model of higher-order themes was created 
to have an overall picture. The research questions were mapped with 
the corresponding themes. Finally, the extracted data were summarized in categories which 
are reported in the results section (in pictures or tables) associated with the research 
questions they belong to. 
\vskip 7mm
\section{Results}
In this section we present the results we found from the selected studies to 
answer each research question. We illustrate the different categories of problems,
techniques, hybridization classes, evaluation methodologies, etc. with examples from
the included studies. The results are further discussed in the next section.  
\vskip 5mm
\subsection{RQ1: Included studies}
\begin{figure}[!htbp]
	\centering
	\includegraphics[width=100mm,scale=0.4]{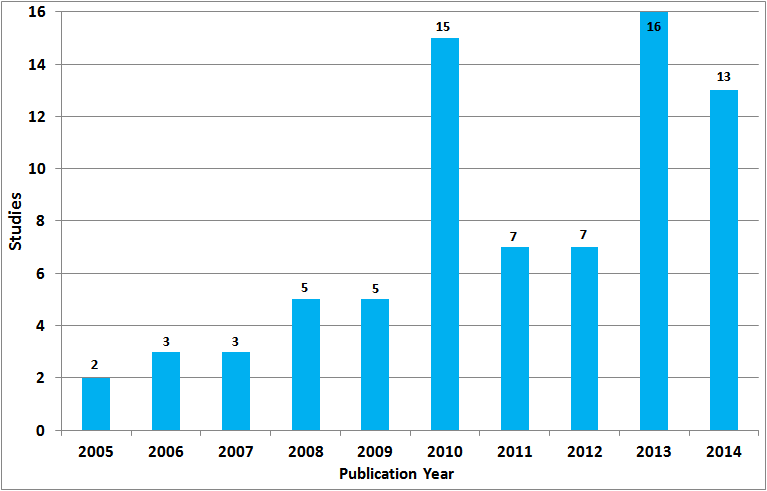}
	\caption{Distribution of studies per publication year}
\end{figure}
\noindent RQ1 addresses the most relevant studies that present Hybrid RSs. We selected 
76 papers as the final ones for further processing. They were published in conference 
proceedings and journals from 2005 to 2015. The publication year distribution of the 
papers is presented in Figure 2. It shows that most of the hybrid RS papers we selected 
were published in the last 5 years.  
\par
For the quality assessment process we used the quality questions listed in Table 5.   
In Figure 3, the box plots of quality score distributions per study type (conference or journal) 
are shown. We see that about 75\% of journal studies have quality score higher than 0.9. Same is 
true for about 35\% of conference studies. In Figure 4 we present the average quality score about 
each quality question.  
\begin{figure}[!htbp]
	\centering
	\includegraphics[width=100mm,scale=0.4]{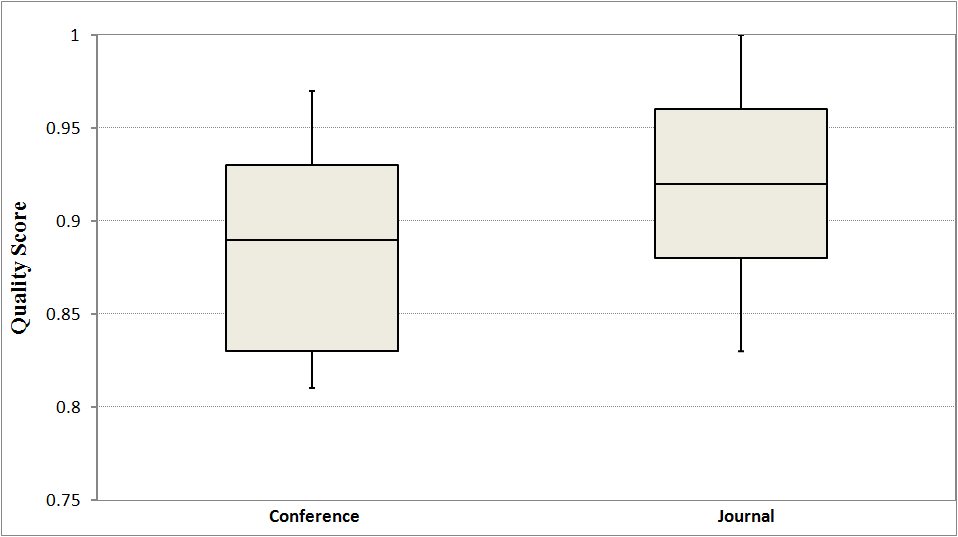}
	\caption{Boxplot of quality score per publication type}
\end{figure} 
\begin{figure}[!htbp]
	\centering
	\includegraphics[width=100mm,scale=0.5]{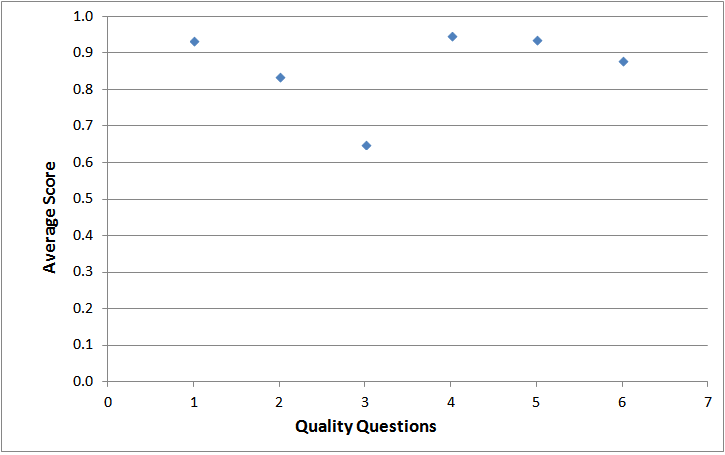}
	\caption{Average score of each quality question}
\end{figure}
QQ4 (Did the study describe the components or architecture of 
the proposed system?) has the highest average score (0.947) wheres QQ3 (Did the study suggest 
further research?) has the lowest (0.651). The weighted quality score is higher than 0.81 for any 
included paper. Only one journal study got a weighted average score of 1.0 (highest possible).  
\vskip 5mm
\subsection{RQ2: Research problems}
To answer RQ2 we summarize the most important RS problems the studies try to solve. A total of 12 
problems were found. The most frequent are presented in Figure 5 with the corresponding number of 
studies where they appear. Studies may (and often do) address more than one problem. 
Same thing applies for other results (data mining techniques, domains, evaluation metrics, etc.) 
reported in this section. Below we describe each of the problems: 
\begin{description}[leftmargin=0.25cm]
	%
	\setlength\itemsep{0em}
	\item[\textbf{Cold-start}] This problem is heavily addressed in the literature 
	\cite{Lika20142065,0295-5075-92-2-28002} and has to do with recommendations for new  
	users or items. In the case of new users the system has no information about their preferences 
	and thus fails to recommend anything to them. In the case of new items the system has no ratings 
	for these items and doesn't know to whom recommend them. To alleviate cold-start, authors in 
	\hyperlink{P21}{[P21]} use a probabilistic model to extract latent features from item's 
	representation. 
	Using the latent features they generate accurate pseudo ratings, even in cold-start situation 
	when few or no ratings are provided. Another example is \hyperlink{P47}{[P47]} where the 
	authors try to solve the new user cold-start in the e-learning domain by combining CF with a 
	CBF representation of learning contents. Cold-start problem is also treated in 
	\hyperlink{P26}{[P26]} where the authors merge the weighted outputs of different recommendation 
	strategies using Ordered Weighted Averaging (OWA), a mathematical technique first introduced 
	in \cite{Yager:1988:OWA:46931.46950}. In total, cold-start was found in 23 studies.
	\item[\textbf{Data sparsity}] This problem rises from the fact that users usually rate a 
	very limited number of the available items, especially when the catalog is very large. 
	The result is a sparse \emph{user-item} rating matrix with insufficient data for identifying 
	similar users or items, negatively impacting the quality of the recommendations. 
	Data sparsity is prevalent in CF RSs which rely on peer feedback to provide recommendations. 
	In \hyperlink{P13}{[P13]} data sparsity of cross-domain recommendations is solved using
	a factorization model of the triadic relation \emph{user-item-domain}. Also in 
	\hyperlink{P1}{[P1]} we find an attempt to solve data sparsity by treating each user-item
	rating as predictor of other missing ratings. They estimate the final ratings by merging 
	ratings of the same item by other users, different item ratings made by the same user and
	ratings of other similar users on other similar items. Another example is 
	\hyperlink{P5}{[P5]} where CF is combined with Naive Bayes in a switching way. 
	Data sparsity was a research problem of 22 studies. 
	\item[\textbf{Accuracy}] Recommendation accuracy is the ability of a RS to correctly predict 
	the item preferences of each user. Much attention has been paid to improve the recommendation 
	accuracy since the dawn of RSs. Obviously there is still place for recommendation accuracy 
	improvements. This is especially true in data sparsity situations, as accuracy and data 
	sparsity are two problems that appear together in 6 studies (e.g., \hyperlink{P24}{[P24]}). 
	In \hyperlink{P51}{[P51]} a Bayesian network model with user nodes, item nodes, and feature
	nodes is used to combine CF with CBF and attain better recommendation quality. Other 
	example is \hyperlink{P53}{[P53]} where a web content RS is constructed. The authors 
	construct user's long term interest based on his/her navigation history. Than the similarity
	of user's profile with website content is computed to decide whether to suggest the website
	or not. Experiments conducted with news websites show improved accuracy results.  
	Improving accuracy was a research objective of 16 studies.   
	\item[\textbf{Scalability}] This is a difficult to attain characteristic which is related to 
	the number of 	users and items the system is designed to work for. A system designed to 
	recommend few items to some hundreds of users will probably fail to recommend hundreds 
	of items to millions of people, unless it is designed to be highly scalable. Hyred in 
	\hyperlink{P28}{[P28]} is an example of a system designed to be scalable and overcome 
	data sparsity problem as well. The authors combine a modified Pearson correlation CF 
	with distance-to-boundary CBF. They find the nearest and furthest neighbors of 
	each user to reduce the dataset. The use of this compressed dataset improves scalability,
	alleviates sparsity, and also slightly reduced the computational time of the system.  
	In \hyperlink{P69}{[P69]} the authors propose a hybrid RS designed to recommend images
	in social networks. They involve CF and CBF in a weighted way and also consider aesthetic 
	characteristics of images for a better filtering, which overcomes the problem of scalability 
	and cold-start as well. In \hyperlink{P29}{[P29]} a system with better scalability is conceived
	by combining Naive Bayer and SVM with CF. Improving scalability was addressed 
	in 11 studies. 
	\item[\textbf{Diversity}] This is a desired characteristic that is getting attention 
	recently \cite{Ge:2010:BAE:1864708.1864761}. Having diverse recommendations is 
	important as it helps to avoid the popularity bias. The latter is having a recommendation
	list with items very similar to each other (e.g., showing all the episodes of a 
	very popular saga). 
	A user that is not interested in one of them is probably not interested in any of them and
	gets no value from that recommendation list. \emph{K-Furthest Neighbors}, the inverted neighborhood model of K-NN is used in \hyperlink{P12}{[P12]} for the purpose of creating 
	more diverse recommendations. The authors report an increased diversity. However, the 
	user study they conduct shows that the perceived usefulness of it is not different from 
	the one of traditional CF. In \hyperlink{P46}{[P46]} the concept of Experts is utilized 
	to find novel and relevant items to recommend. The ratings of users are analyzed and 
	some of the users are promoted as "experts" of a certain taste. They generate 
	recommendations of their for the rest of the "normal" users in that item taste. 
	Diversity is also addressed in \hyperlink{P36}{[P36]} totaling in 3 studies.
	\item[\textbf{Other}] These are other problems appearing in few studies. They include 
	Lack of Personalization, Privacy Preserving, Noise Reduction, 
	Data source Integration, Lack of Novelty and User preference Adaptiveness. 
\end{description}  
\begin{figure}[!htbp]
	\centering
	\includegraphics[width=100mm,scale=0.5]{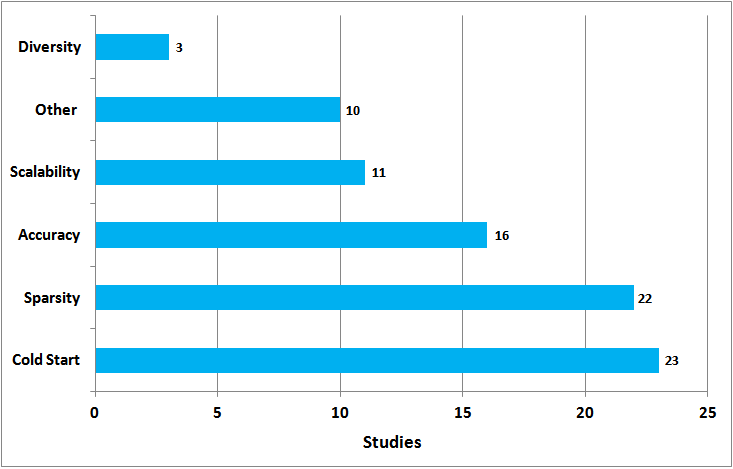}
	\caption{Addressed problems}
\end{figure}
\vskip 5mm
\subsection{RQ3a: Data mining and machine learning techniques}
In this section we address the distribution of the studies according to the basic
Data Mining (DM) and Machine Learning (ML) techniques they use to build their hybrid RSs. 
The variety of DM and ML techniques or algorithms used is high.  
Authors typically use different techniques to build the diverse components of their 
solutions or prototypes. In Table 7 we present the most frequent that were found in the 
included studies. Below we describe some of them. More details about the characteristics of 
DM/ML techniques and how they are utilized to build RSs can be found at \cite{Amatriain2011}. 		
\begin{description}[leftmargin=0.25cm]
	%
	\setlength\itemsep{0em}
	\item[\textbf{K-NN}] \emph{K-Nearest Neighbors} is a well known classification 
	algorithm with several versions and implementations, widely utilized in numerous 
	data mining and other applications. This technique is 
	popular among collaborative filtering RSs which represent the most common family 
	of recommenders. It is mostly utilized to analyze neighborhood and find users of 
	similar profiles or analyze items' catalog and find items with similar characteristics. 
    K-NN was found in a total of 59 studies. 
	\item[\textbf{Clustering}] There are various clustering algorithms used in RSs and other 
	data mining applications. They typically try to put up a set of categories with which 
	data can be identified. The most popular is \emph{K-means} which partitions the entire data 
	into K clusters. In RSs clustering is mostly applied to preprocess the data. 
	In \hyperlink{P6}{[P6]} the authors experiment with K-way (similar to K-means) clustering
	and \emph{Bisecting K-means} for grouping different types of learning items. They also 
	use CBF to create learners' profiles and build an e-learning recommender with improved 
	accuracy. An other example is \hyperlink{P44}{[P44]} where websites are clustered using 
	co-occurence of pages and the content data of pages. The results are aggregated to get 
	the final recommendations and overcome data sparsity. In total clustering algorithms 
	were used in 34 studies.
	\item[\textbf{Association rules}] Association rule mining tries to discover valuable relations 
	(association rules) in large databases of data. These associations are in the form 
	X \verb|=>| Y, where X and Y are sets of items. The association that are above a 
	minimum level of support with an acceptable level of confidence can be used to derive 
	certain conclusions. In recommender systems this conclusions are of the form "X likes Y" 
	where X is a user to whom the system can recommend item Y. In \hyperlink{P58}{[P58]}
	information collected from a discussion group is mined and association rules are used 
	to form the user similarity neighborhood. Word Sense Disambiguation is also used to 
	select the appropriate semantically related concept from posts which are then recommended
	to the appropriate users of the forum. This hybrid meliorates different problems such as
	cold-start, data sparsity and scalability. In \hyperlink{P59}{[P59]} classification 
	based on association methods is applied to build a RS in the domain of tourism. The 
	system is more resistant to cold-start and sparsity problems. To overcome cold-start,
	the authors in \hyperlink{P61}{[P61]} propose a procedure for finding similar items by 
	association rules. Their algorithm considers the user-item matrix as a transaction 
	database where the user Id is the transactional Id. They find the support of each item
	and keep items with support greater than a threshold. Afterwards, they calculate the 
	confidence of remaining rules and rule scores by which they find the most similar 
	item to any of the items. Association rules were found in 17 studies.
	\item[\textbf{Fuzzy logic}] Also called \emph{fuzzy set theory} it is a set 
	of mathematical 
	methods that can be used to build hybrid RSs. Those methods are also called 
	reclusive in the literature. Contrary to CF which relies on neighborhood preferences 
	without considering item characteristics, they require some representation of the 
	recommended items \cite{Yager:2003:FLM:794087.794088}. Reclusive methods are complementary 
	to collaborative methods and are often combined with them to form hybrid RSs. 
	An example of using Fuzzy logic is \hyperlink{P27}{[P27]} where better accuracy is 
	achieved by combining 2 CFs with a fuzzy inference system in a weighted way to 
	recommend leaning web resources. In \hyperlink{P34}{[P34]} fuzzy clustering is used to 
	integrate user profiles retrieved by a CF with Point Of Interest (POI) data retrieved
	from a context aware recommender. The system is used in the domain of tourism and 
	provides improved accuracy. In total Fuzzy logic was found in 14 studies. 
	\item[\textbf{Matrix manipulation}] Here we put together the different methods and 
	algorithms that are based on matrix operations. The methods we identified are 
	Singular Value Decomposition (SVD), Latent Dirichlet Allocation (LDA), 
	Principal Component Analysis (PCA), Dimensionality Reduction and 
	similar matrix factorization techniques. Matrix manipulation methods are often used 
	to build low error collaborative RSs and were especially promoted after the Netflix challenge 
	was launched in 2006. In \hyperlink{P75}{[P75]} a topic model based on LDA is used to 
	learn the probability that a user rates an item. An other example is \hyperlink{P76}{[P76]} 
	where Dimensionality Reduction is used to solve sparsity and scalability in a 
	multi-criteria CF. They were found in 9 studies.
	\item[\textbf{Other}] Other less frequent techniques such as Genetic Algorithms, 
	Naive Bayes, Neural Networks, Notion of Experts, Statistical Modeling, etc. were found 
	in 19 papers.  
\end{description}  		
\begin{table}[ht] 
	\caption{Distribution of studies by DM/ML techniques} 
	\centering 
	\small 
	\begin{tabular}
		{l c}  
		\topline 
		\headcol \textbf{DM/ML technique \qquad\quad } & \textbf{Studies}  \\ [0.5ex] 
		\midline   
		K-NN & 59       \\    
		\rowcol Clustering & 34						  	\\
		Association rules & 17			 				 \\
		\rowcol Fuzzy logic & 14						 \\
		Matrix manipulation & 9			  				\\
		\rowcol Other & 19						 		\\
		\bottomline
	\end{tabular}  
\end{table}
\vskip 5mm
\subsection{RQ3b: Recommendation technique combinations}
In this section we present a list of the most common technique combinations 
that form hybrid RSs. We also present the problems each of this combinations is 
most frequently associated with. In the following subsections the construct and 
technical details of some of the prototypes implementing each combination is 
described. Table 8 presents the summarized results.
\begin{table}[ht]
	\caption{Hybrid recommendation approaches distributed per problem}
	\small 
	\centering      
	\small
	\begin{tabular}{l c c c c c c}
		\topline
		\headcol & \multicolumn{6}{c}{\textbf{Hybrid recommenders and studies}} \\
		\cline{2-7} 
		\headcol \textbf{Problem} & \textbf{CF-X} & \textbf{CF-CBF} & \textbf{CF-CBF-X} 
		& \textbf{IICF-UUCF} & \textbf{CBF-X} & \textbf{Other} \\
		\midline
		\multicolumn{1}{l}{Cold-start} & 2 & 3 & 2 & 1 & 1 & 5  \\
		\rowcol \multicolumn{1}{l}{Data Sparsity} & 0 & 5 & 3 & 3 & 4 & 6  \\
		\multicolumn{1}{l}{Accuracy} & 2 & 3 & 0 & 2 & 2 & 4  \\
		\rowcol \multicolumn{1}{l}{Scalability} & 0 & 2 & 2 & 0 & 2 & 2  \\
		\multicolumn{1}{l}{Diversity} & 2 & 0 & 0 & 0 & 0 & 1  \\
		\rowcol \multicolumn{1}{l}{Other} & 0 & 2 & 1 & 1 & 1 & 2  \\
		\hline
		\multicolumn{1}{l}{\textbf{Total}} & \textbf{6} & \textbf{15} & \textbf{8} 
		& \textbf{7} & \textbf{10} & \textbf{20}  \\
		\bottomline
	\end{tabular}
\end{table} 
\subsubsection{CF-X}
Here we report studies that combine CF with one other technique which is not CBF 
(those are counted as CF-CBF). 
An example of this combination is \hyperlink{P8}{[P8]} where the authors go hybrid
to improve the performance of a multi-criteria recommender. They base their solution on 
the assumption that usually only a few selection criteria are the ones which 
impact user preferences about items and their corresponding ratings. Clustering is 
used first to group users based on the items' criteria they prefer. CF is then used 
within each cluster of similar users to predict the ratings. They illustrate their 
method by recommending hotels from 
TripAdvisor\footnote{\url{http://www.tripadvisor.co.uk}} and report performance 
improvements over traditional CF. 
Other attempt to improve the predictive accuracy of traditional CF is 
\hyperlink{P60}{[P60]}. Here the authors integrate in CF discrete demographic 
data about the users such as \emph{gender}, \emph{age}, \emph{occupation}, etc.  
Fuzzy logic is used to compute similarities between users utilizing this extra demographic 
data and integrate the extra similarities with the user-based similarities calculated 
from ratings history. After calculating the final user similarities their 
algorithm predicts the rating values. The extra performance which is gained from the 
better user similarities that are obtained, comes at the cost of a slightly larger 
computational time which is however acceptable.  
In total CF-X combination was found in 6 studies with X being KBF, DF or a DM/ML 
technique from those listed in Table 6.
\subsubsection{CF-CBF}
This is a very popular hybrid RS utilizing the two most successful recommendation 
strategies. In many cases the recommendations of both systems are weighted to produce 
the final list of predictions. In other cases the hybrid RS switches from CF to 
CBF or is made up of a more complex type of combination (see section 3.5). 
An example is \hyperlink{P28}{[P28]} where the authors develop a hybrid RS suitable 
for working with high volumes of data and solve scalability problems in e-commerce systems. 
Their solution first involves CF (Pearson's product moment coefficients) to reduce the 
dataset by finding the nearest neighbors of each user, discarding the rest and reducing the 
dataset. Afterwards distance-to-boundary CBF is used to define the decision boundary 
of items purchased by the target user. The final step combines the CF score (correlation 
coeficient between two customers) with the distance-to-boundary score (distance between 
the decision boundary and each item) in a weighted linear form. The authors report an 
improved accuracy of their hybrid RS working in the reduced dataset, compared to other 
existing algorithms that use full datasets. 
\par
In \hyperlink{P51}{[P51]} the authors propose a CF-CBF hybrid recommender which is based 
on Bayesian networks. This model they build uses probabilistic reasoning to compute the 
probability distribution over the expected rating. The weight of each recommending 
strategy (CF and CBF) is automatically selected, adapting the model to the 
specific conditions of the problem (it can be applied to various domains). The authors
demonstrate that their combination of CF and CBF improves the recommendation accuracy.  
Other studies involve similar mathematical models or constructs (e.g., fuzzy logic)
to put together CF and CBF and gain performance or other benefits. In total CF-CBF 
contributions were found in 15 studies. 
\subsubsection{CF-CBF-X}
Those are cases in which CF and CBF are combined together with a third approach. 
One example is \hyperlink{P14}{[P14]} where CF and CBF are combined with DF to generate 
recommendations for groups of similar profiles (users). These kind of recommendations are 
particularly useful in online social networks (e.g., for advertising). The goal of 
the authors is to provide good recommendations in data sparsity situations. First CBF is 
used to analyse ratings and items' attributes. CF is then invoked as the second stage 
of the cascade to generate the group recommendations. DF is used to reinforce CF in the 
cases of sparse profiles (users with few ratings). In total CF-CBF-X was found in 8 
studies. X is mostly a clustering technique or DF. 
\subsubsection{IICF-UUCF}
Item-Item CF and User-User CF are two forms of CF recommenders, differing on the way 
the neighborhoods are formed. Some studies combine both of them to improve overall 
CF performance. 
An example is \hyperlink{P70}{[P70]} where the authors present a hybrid recommendation 
framework they call Collaborative Filtering Topic Model (CFTM) which considers both 
user's reviews and ratings about items of a certain topic (or domain) in e-commerce. 
The first stage which is offline performs sentiment analysis in the reviews to calculate 
the User or Item similarity. The second stage of the cascade uses IICF or UUCF (switching) 
to predict the ratings. The authors evaluate using 6 datasets of different domains from Amazon 
and report that their hybrid approach performs better than traditional CF, especially 
in sparsity situations. IICF-UUCF combinations were found in 7 studies. 
\subsubsection{CBF-X}
There were also 10 studies in which CBF is combined with another technique X which is 
not CF (counted as CF-CBF). X represents different approaches like KBF and DF or 
DM/ML techniques like clustering etc. One example is   
\hyperlink{P63}{[P63]} where the authors describe and use the interesting notion of 
\emph{user lifestyle}. They select demographic information, consumer credit data and 
TV program preferences as lifestyle indicators, and confirm their significance by performing 
statistical analysis on 502 users. The most significant lifestyle attributes are binary 
encoded and used to form the neighborhoods and ratings of each user by means of Pearson 
correlation. The authors call the resulting complete (in terms of ratings) matrix
\emph{pseudoUser - item} matrix. It is then used for a Pearson based (classical CF)
prediction of the original \emph{user-item} ratings. Considerable performance 
improvements are reported. 
\subsubsection{Other}
Other implementations include combinations of the 
same recommendation strategy (e.g., \emph{CF1-CF2} with different similarity measures 
or tuning parameters each), trust-aware recommenders that are being used in social 
communities, prototypes using association rules mining, neural networks, genetic algorithms, 
dimensionality reduction, social tagging, semantic ontologies, pattern mining or 
different machine learning classifiers.	  
\vskip 5mm
\subsection{RQ4: Classes of hybridization}
To answer RQ4 we classified the examined hybrid RSs according to the taxonomy proposed 
by Burke \cite{Burke:2002:HRS:586321.586352}. This taxonomy categorizes hybrid RSs in 7 
classes based on the way the different recommendations techniques are aggregated with each 
other. Each class is explained in the subsections below where we discuss in more details 
few examples from the included papers. The results are summarized in Figure 6.
\begin{figure}[!htbp]
	\centering
	\includegraphics[width=100mm,scale=0.5]{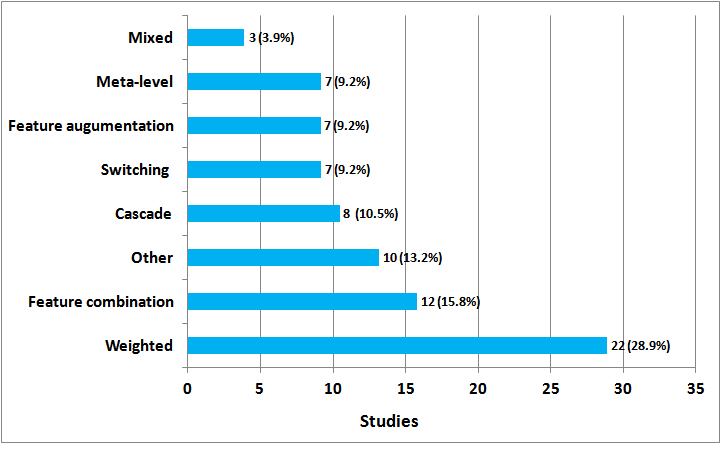}
	\caption{Distribution of studies per hybridization class}
\end{figure} 
\subsubsection{Weighted}
Weighted hybrids were the most frequent. They compute the scores of the items they 
recommend by aggregating the output scores of each recommendation technique 
using weighted linear functions. 
One of the first weighted recommenders was P-Tango \cite{DECAMPOS2010785} 
which combined CF and CBF rating scores in a linear weighted way to recommend online 
newspapers. In P-Tango, aggregation was made giving equal initial weights to each score 
and then possibly adapting by the feedback of users. The weights of CF and CBF are 
set on a per-user basis enabling the system to determine the optimal mix for each user and 
alleviating the "gray sheep" problem. 
In \hyperlink{P38}{[P38]} the authors propose a weighting method for combining
user-user, user-tag and user-item CF relations in social media. The method they propose 
computes the final rating score of an item for a user as the linear combination of the 
above three CF relations. Unlike the traditional CF, this weighted hybrid CF recommender 
is completely based on tags and does not require that users provide explicit rating scores 
for the items that are recommended (e.g., photos). 
An other example is \hyperlink{P6}{[P6]} where the authors combine a content-based model 
with a rule-based model to recommend e-learning materials. They build 
their CBF using an education domain ontology and compute the scores 
of each learning material using \emph{Vector Space Model} and \emph{TF-IDF}. The 
rule-based recommender utilizes the ontology and the user's previously visited concepts 
to realize a semantic mapping between user's query and his/her semantic profile, 
resulting in adequate term recommendations about learning materials. The two RS modules 
set different weights to each recommended item based on user's preferences and higher 
accuracy is achieved. 
Apparently the benefit of a weighted hybrid is the fact that it uses a straightforward way to 
combine the results of each involved technique. It is also easy to adjust priority 
assignment for each involved strategy by changing the weights. This class of hybrid 
RS was used in 22 (28.9\%) of the included studies.
\subsubsection{Feature combination}
This type of hybrid RSs treats one recommender's output as additional feature data, 
and uses the other recommender (usually content-based which makes extensive use of
item features) over the new extended data. In case of a CF-CBF hybrid, the system
does not exclusively rely on the collaborative data output of CF. That output is 
considered as additional data for the CBF which generates the final list. This
reduces the sensitivity to possible sparsity of the initial data. 
For example, in \hyperlink{P40}{[P40]} the authors present a CF-CBF book recommender which 
implements an extended feature combination strategy. In the first phase new features 
(prefered books) are generated by applying CF among the readers. In the second phase 
they utilize \emph{fuzzy c-means clustering} and \emph{type-2 fuzzy logic} to obtained 
data for creating book categories of each user type (teacher, researcher, student). 
In the third and final phase CBF is involved to recommend the most relevant books to 
each user. The authors report performance improvements both in MAE and 
F1 accuracy scores.  
Also in \hyperlink{P25}{[P25]} the authors build an information system about courses and 
study materials for scholars. The system invokes a web crawler to collect related web 
pages and classifies the obtained results in different item categories (websites, 
courses, academic activities) using a web page classifier supported by a school 
ontology. An information extractor is later invoked to get significant web page features. 
Finally the system operates on the extra features of each item category to produce 
integrated recommendations based on the order of the keyword weight of each item.
System verification reports higher recommendation quality and reliability. 
Feature combination hybrids were found in 12 (15.8\%) studies.
\subsubsection{Cascade}
Cascade hybrids are examples of a staged recommendation process. First one technique 
is employed to generate a coarse ranking of candidate items and than a second technique 
refines the list from the preliminary candidate set. Cascades are order-sensitive; 
A CF-CBF would certainly produce different results from a CBF-CF.  
An example is \hyperlink{P67}{[P67]} which presents a mobile music cascade recommender 
combining SVM genre classification with collaborative user personality diagnosis. 
The first level of the recommendation process consists of a multi-class SVM classifier
of songs based on their genre. The second level is a personality diagnosis which assumes
that user preferences for songs constitute a characterization of their underlying 
personality. The personality type of each user is assumed to be the vector of 
ratings in the items the user has seen. The personality diagnosis approach estimates the
probability that each active user is of the same personality type as other users. As a 
result the probability that a active user will like new songs is computed in a more
personalized way.  
\par
In \hyperlink{P49}{[P49]} the authors combine two CF systems with different properties. 
The first module is responsible for retrieving the data and generating the list of 
neighbors for each user. This module uses two distance measures, Pearson's coefficient 
and Euclidean distance in a switching way, depending on the user's deviation from 
his/her average rating. The authors report that Euclidean distance performs better 
than Pearson's coefficient in most of the cases. In the second module of the cascade, 
they experiment switching between three predictors to generate the final 
recommendations: Bayesian estimator, Pearson's weighted sum and adjusted weighted 
sum. They also report that the Bayesian prediction gives best results.
An other example of a cascade hybrid is \hyperlink{P68}{[P68]}. It implements a 
cascade of item-based CF and Sequential Pattern Mining (SPM) to recommend items 
in an e-learning environment. To adopt the CF to the e-learning domain they introduce 
a damping function which decreases the importance of "old" ratings. The SPM 
module takes in a list of k most similar items for each item and determines it support.
At the end it prunes the items with support less than the 
threshold and generates the recommended items. The authors also apply this 
recommender in P2P learning environments for resource pre-fetching. Cascade 
hybrids were found in 8 (10.5\%) studies.
\subsubsection{Switching}
In a switching hybrid the system switches between different recommendation 
techniques according to some criteria. For example, a CF-CBF approach can 
switch to the content-based recommender only when the collaborative strategy 
doesn't provide enough credible recommendations. Even different versions of 
the same basic strategy (e.g., CBF1-CBF2 or CF1-CF2) can be integrated in a 
switching form.
An example is DailyLearner, an online news recommender presented in 
\cite{Billsus:2000:LAW:325737.325768}. It first employs a short-term CBF recommender 
which considers the recently rated news stories utilizing \emph{Nearest Neighbor} 
text classification and \emph{Vector Space Model} with \emph{TF-IDF} weights. 
If a new story has no near neighbors the system switches to the long-term model
which is based on data collected over a longer time period, presenting user's 
general preferences. It uses a Naive Bayes classifier to estimate the probability 
of news being important or not. 
\par
In \hyperlink{P29}{[P29]} the authors build a switching hybrid RS that is based 
on a Naive Bayes classifier and Item-Item CF. The classifier is trained in 
offline phase and used to generate the recommendations. If this recommendations 
have poor confidence the Item-Item CF recommendations are used instead. First, 
they compute the posterior probability of each class generated by the Naive 
Bayes classifier. Then they assume that the classifier's confidence is high if 
the posterior probability of the predicted class is sufficiently higher than the 
ones of the other classes. Movielens and Filmtrust are employed to evaluate 
the approach and performance improvements are reported, both in accuracy and 
in coverage. 
An other example of a switching hybrid is \hyperlink{P55}{[P55]} where the authors 
describe the design and implementation of a mobile locaton-aware CF-KBF recommender 
of touristic sites (e.g., restaurants). Their system involves both
CF and KBF modules in generating recommendations. Then 3D-GIS location data are used to 
compute the physical distance of the mobile user from the recommended sites. The system
switches from one recommendation strategy to the other and performs a distance-based 
re-ranking of the recommendations, choosing the sites that are physically closer to the 
user with higher accuracy.  
In most of the cases we see that complexity of switching RSs lies in the switching 
criteria which are mostly based on distance or similarity measures. However, this systems 
are sensitive to the strengths and weaknesses of the composing techniques. This hybrid 
RS category was found in 7 (9.2\%) studies. 
\subsubsection{Feature augmentation}
In this class of hybrids, one of the combined techniques is used to 
produce an item prediction or classification which is then comprised in the operation of the 
other recommendation technique. Feature augmentation hybrids are order-sensitive as the 
second technique is based on the output of the first. For example an association rules 
engine can generate for any item, similar items which can be used as augmented item 
attributes inside a second recommender to improve its recommendations. Libra presented in \cite{Mooney:2000:CBR:336597.336662} 
is a content-based book recommender. It augments the textual features of the books with
"related authors" and "related titles" data obtained from Amazon CF recommender to obtain
a better recommendation quality. Libra uses an inductive learner to create user profiles. 
This inductive learner is based on vectorized bag-of-words naive Bayes text classifier.
The authors report that the integrated collaborative content has a significant positive 
effect on recommendation performance. 
\par
\hyperlink{P36}{[P36]} presents a hybrid method which combines multidimensional clustering 
and CF to increase recommendation diversity. They first invoke multidimensional clustering to 
collect and cluster user and item data. Clusters with similar features are deleted and the 
remaining feature clusters are fed into the CF module. Item-Item similarity is computed 
using an adjusted cosine similarity which works for \emph{m} cluster features of each 
item. Finally the rating predictions are computed base on item-item similarity and the 
rating deviations from neighbors. The authors report an increase in recommendation 
diversity with only minimal loss in accuracy. 
Feature augmentation offers a means of improving the performance of a system (in the
above examples the second recommender) without the need to modify it. The extra 
functionality is added by augmenting the processed data. This hybrid RS class was 
used in 7 (9.2\%) studies.
\subsubsection{Meta level}
Meta levels are also an example of order-sensitive hybrid RSs that use an entire model 
produced by the first technique as input for the second technique. 
It is typical to use content-based recommenders to build item representation models, and then 
employ this models in collaborative recommenders to match the items with user profiles. 
A meta level recommendation strategy was implemented by Fab 
\cite{Balabanovic:1997:FCC:245108.245124}, one of the first website recommenders. 
Fab uses a selection agent which based on \emph{term vector model} accumulate user-specific 
feedback about areas of interest for each user. There are also two collection agents: 
search agents which perform a search for websites, and index agents which construct queries 
for already found websites to avoid duplicate work. Collection agents utilize the models 
of the users (collaborative component) to collect the most relevant websites which are 
then recommended to the users. 
\par
Also \hyperlink{P20}{[P20]} presents a meta level recommender used in the domain 
of music which integrates CF with CBF. Here each user is stochastically matched with a 
music genre based on the collaborative output. Then the system generates a musical piece 
for the user based on the acoustic features. For the integration they adopt 
a probabilistic generative model called \emph{three-way aspect model}. As this model is 
only used for textual analysis and indexing (bag-of-words representation) they propose 
the \emph{bag-of-timbres} model, an interesting approach to content-based music 
recommendations which represents each musical piece as a set of polyphonic timbres. 
The advantage this hybridization class presents is that the learned model of the first 
technique is compressed and thus better used from the second. However, the integration 
effort is considerable and use of advanced constructs is often required. This hybrid 
RS class was found in 7 (9.2\%) studies.
\subsubsection{Mixed}
Mixed hybrids represent the simplest form of hybridization and are reasonable when it is 
possible to put together a high number of different recommenders simultaneously. Here the 
generated item lists of each technique are added to produce a final list of 
recommended items. 
One of the first examples of mixed hybrids was PTV system \cite{Smyth200053} which used 
CBF to relate similar programs to the user profile and CF to relate similar user profiles together. 
The CBF module converts each user profile in a feature-based representation they call
\emph{profile schema} which is basically a TV program content summary represented in 
features. The CF module computes the similarity of two users utilizing a graded difference 
metric of the ranked TV programs in each user's profile. At the end, a selection of programs
recommended by the two modules is suggested.
\par
Yet another example of recommending TV programs is a CF-CBF mixed hybrid named queveo.tv 
described in \hyperlink{P52}{[P52]}. Here the authors use demographic information such as 
age, gender and profession together with user's history to build his/her profile which is used 
by the CBF module. This module makes use of Vector Space Model and cosine correlation to 
provide the recommended TV programs. The CF module uses both user-based CF to generate
the top neighbors of the active user, and item-based CF to predict the level of interest 
of the user for a certain item. At the end the system takes recommendations from the two 
modules to generates the final list of TV programs. Those TV programs that were part of
both listings (CBF and CF) are highlighted as \emph{Star Recommendations}, as they are 
probably the most interesting for the user. 
Mixed hybrid RSs are simple and can eliminate acute problems like cold-start 
(new user or new item). They were found in 3 (3.9\%) studies only. 		
\vskip 5mm
\subsection{RQ5: Application domains}
A rich collection of 18 application domains was identified. Figure 7 presents the percentage 
of studies for each application domain. We see that most of the studies (21 or 27.6\%) 
are domain independent. They haven't been applied to a particular domain. Movie domain was 
considered by 17 (22.3\%) studies. Next comes education or e-learning considered by 
9 (11.8\%) studies. 
\begin{figure}[!htbp]
	\centering
	\includegraphics[width=100mm,scale=0.5]{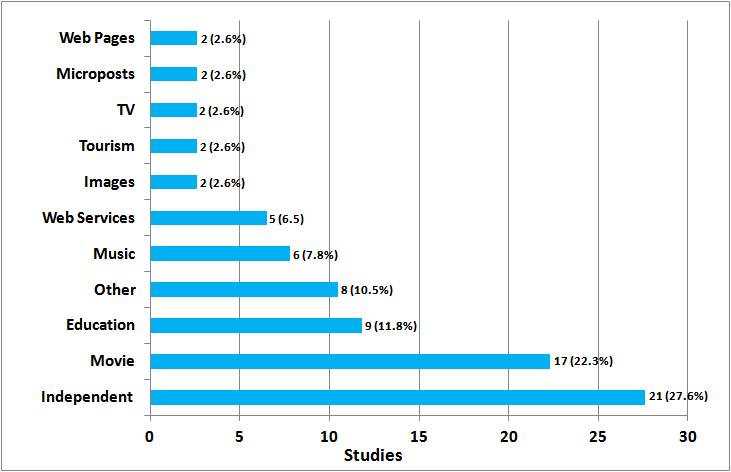}
	\caption{Distribution of studies according to the application domains}
\end{figure}
Six (7.8\%) studies were applied in the domain of music. There were also web service RSs 
implemented in 5 (6.5\%) studies. Other domains are images, touristic sites, TV programs, 
web pages and microposts which appeared in 2 (2.6\%) studies each. Domains like business, 
food, news, bibliography, etc. categorized as "Other" count for less than 10.5\% of the 
total number of studies.  
\vskip 5mm
\subsection{RQ6: Evaluation}
Another important aspect of hybrid RSs that we examined is the evaluation process. 
In this section we present results about the evaluation methodologies and the corresponding 
involved metrics (answering RQ6a), evaluated RS characteristics and the utilized 
metrics for each (answering RQ6b) and finally the public datasets used to train and test 
the algorithms (answering RQ6c). 
\subsubsection{RQ6a: Evaluation Methodologies}
\begin{table}[ht] 
	\caption{Evaluation methodology} 
	\small 
	\centering      
	\begin{tabular}
		{l c}  
		\topline 
		\headcol \textbf{Methodology} & \textbf{Studies}  	\\ [0.5ex] 
		\midline   
		Comparison with similar method & 58       			\\   
		\rowcol User survey & 14						  	\\
		Comparison and user survey & 3			  			\\
		\rowcol No evaluation & 1						  	\\
		\bottomline
	\end{tabular}  
\end{table}
\noindent Here we try to explain how (with what methodologies) the evaluation process 
is performed and what metrics are involved in each methodology.  Table 9 lists the 
distribution of studies according to the methodology they use to perform the evaluation. 
There are 58 (more than three-quarters) studies comparing the proposed 
system (or solution) with a similar well known method or technique. Usually CF-X or CF-CBF 
hybrid RSs are compared with pure CF or CBF. In some cases the proposed system is 
compared with different parameter configurations of itself. 
Accuracy or error measures like MAE (Mean Average Error) or RMSE (Root Mean Square Error) 
are very common. They estimate the divergence of the RS predictions from the actual ratings. 
Decision support metrics like Precision, Recall and F1 are also very frequent. Precision is 
the percentage of selected items that are relevant. Recall is the percentage of relevant 
items that are recommended. F1 is the harmonic mean of the two.
User surveys are the other evaluation methodology utilized in 14 studies. 
They mainly perform subjective quality assessment of the RS and require the involvement 
of users who provide feedback for their perception about the system. Surveys are usually 
question based and reflect the opinion of users about different aspects of the 
hybrid recommender. An example of user surveys is \hyperlink{P27}{[P27]} where the 
participants were 30 high school students. In \hyperlink{P50}{[P50]} the users of the 
survey are customers of a web retail store who rated products they purchased. 
In \hyperlink{P74}{[P74]} a mix of real and simulated users are used to rate movies, 
books, etc. In total user surveys were conducted in 14 studies. 
\par
Both comparisons and surveys are used in 3 studies: \hyperlink{P9}{[P9]} where the 
participants were 17 males along with 15 females and different versions of the system were 
compared with each-other, \hyperlink{P12}{[P12]} where the system was compared with 
CF using Movielens and the survey involved 132 participants, and \hyperlink{P40}{[P40]} 
where online user profiles were utilized for the survey, and the proposed fuzzy hybrid book 
RS was compared with traditional CF. The only study with no evaluation at all was 
\hyperlink{P23}{[P23]}. Here the authors present a personalized hybrid recommendation 
framework which integrates trust-based filtering with multi-criteria CF. This framework 
is specifically designed for various Government-to-Business e-service recommendations. 
The authors leave the evaluation of their framework as a future work. 
\subsubsection{RQ6b: Characteristics and metrics}
In order to address RQ6b we analyzed the recommendation characteristics the authors 
evaluate, and what metrics they utilize. Five characteristics were identified, listed 
in Table 10. The top characteristic is accuracy measured in 62 studies. It is 
followed by user satisfaction, a subjective characteristic assessed in 10 studies. 
Diversity is about having different list of recommended items each time the user 
interacts with the system. In total it was measured in 7 studies. Computational 
complexity of the RS is measured in 6 studies. Novelty and serendipity express the 
capability of the hybrid RS to recommend new or even unexpected but still relevant 
items to the user. They were measured in 4 studies.
\begin{table}[ht] 
	\caption{Evaluated characteristics}  
	\small 
	\centering      
	\begin{tabular}
		{l c}  
		\topline
		\headcol \textbf{Recommendation characteristic} & \textbf{Studies}  	\\ [0.5ex] 
		\midline   
		Accuracy & 62       									\\    
		\rowcol User satisfaction & 10			  				\\
		Diversity & 7						  					\\
		\rowcol	Computational complexity & 6			  		\\
		Novelty-Serendipity & 4			  						\\
		\bottomline
	\end{tabular} 
\end{table}
We also observed the metrics that authors use for each evaluated characteristic, summarized
in Table 11. Accuracy is mostly measured by means of precision (31 studies), recall 
(23) and F1 (14). MAE and RMSE were found in 27 and 6 studies 
correspondingly. Other less frequent metrics used to evaluate accuracy include MSE 
(Mean Squared Error), nDCG (normalized Discounted Cumulative Gain), AUC, etc. They were 
found in 15 studies. As previously mentioned user satisfaction is measured by means of 
user surveys which were found in 10 studies. They usually consist of polls which aim to 
get the opinion of the users about different recommendation aspects of the system. 
Diversity is measured mostly by coverage which was found in 4 studies. In the other cases it 
is measured using ranking distances (3 studies). Execution time is the time it takes for the 
system to provide the recommendations and is a measure of the computational complexity. It 
was found in 6 studies. Novelty and Serendipity are measured by less known metrics such as
Surprisal, Coverage in Long-Tail or Expected Popularity Complement.  
\begin{table}[ht] 
	\caption{Evaluated characteristics and involved metrics} 
	\small
	\centering       
	\begin{tabular}
		{l l c}   
		\topline
		\headcol \textbf{Characteristic}  & \textbf{Metrics} & \textbf{Studies}  	\\ [0.5ex]    
		\midline
		Accuracy & Precision & 31       				\\    
		\rowcol & MAE & 27 								\\	
		& Recall & 23         					\\ 
		\rowcol & F1 & 14        								\\ 
		& RMSE & 6										\\ 
		\rowcol & Other & 15							\\
		\hline          
		User satisfaction & Qualitative Subjective Assessment & 10 		\\
		\hline  
		\rowcol Diversity & Coverage & 4 						\\
		& Ranking distances & 3							\\
		\hline 
		\rowcol Complexity & Execution time & 6 				\\	
		\hline          
		Novelty-Serendipity & Surprisal & 2				\\
		\rowcol & Coverage in Long-Tail & 1 					\\
		& Expected Popularity Complement & 1			\\	
		\hline     
	\end{tabular}   
\end{table}
\subsubsection{RQ6c: Datasets} 
We also kept track of the public datasets used by the authors to evaluate their hybrid 
RSs. These datasets are used by the scientific community to replicate experiments 
and validate or improve their techniques. 
\begin{figure}[!htbp]
	\centering
	\includegraphics[width=100mm,scale=0.5]{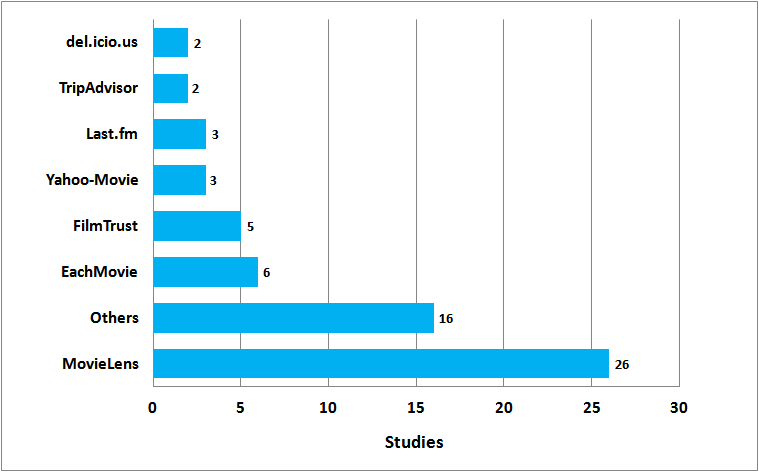}
	\caption{Distribution of studies according to the datasets they use for evaluation}
\end{figure}
There are 55 studies that use at least one public dataset. Sometimes a study uses 
more than one dataset. On the other hand 21 studies do not use any dataset. Sometimes 
they use synthetic data or rely on user surveys or other techniques. In Figure 8 we 
present the datasets that were used and the number of studies in which they appear. 
\begin{description}[leftmargin=0.25cm]
	%
	\setlength\itemsep{0em}
	\item[\textbf{MovieLens}]\footnote{\url{http://grouplens.org/node/73}} used in 26 studies, 
	is one of the most popular public datasets used in the field of RSs. It was collected 
	and made available by GroupLens\footnote{\url{http://grouplens.org}} which is still 
	maintaining it.  
	\item[\textbf{EachMovie}] is also a movie dataset used in 6 studies. Even though it 
	is now retired,  it was the original basis for MovieLens and has been extensively 
	used by the RS community. 
	\item[\textbf{FilmTrust}] is a movie dataset and a recommendation website that uses 
	the concept of trust to recommend movies. It is smaller in size compared to the other 
	movie datasets but it has the advantage of being more recent in content. FilmTrust 
	was used in 5 studies. 		
	\item[\textbf{Yahoo-Movie}] is a dataset containing a subset of Yahoo Movie community
	preferences for movies. It also contains descriptive information about many movies released 
	prior to November 2003. Yahoo-Movie was used in 3 studies.		
	\item[\textbf{Last.fm}]\footnote{\url{http://ocelma.net/MusicRecommendationDataset/lastfm-360K.html}} 
	is a music dataset crawled by last.fm website. It contains information about some of the users' 
	attributes, their track preferences and the artists. Last.fm was used in 3 studies. 		
	\item[\textbf{Tripadvisor}] is a dataset consisting of hotel and site reviews crawled by tripadvisor 
	website. It is especially used to provide touristic recommendations to mobile users.  
	Tripadvisor was used in 2 studies.	
	\item[\textbf{Delicious}]\footnote{\url{http://disi.unitn.it/~knowdive/dataset/delicious/}} 
	is a dataset containing website bookmarks and tags of the form (user, tag, bookmark)
	shared by many users within the network. Delicious dataset was used in 2 studies. 
	\item[\textbf{Other}] less popular datasets containing different type of recommendable 
	items were found in 16 studies.
\end{description}
\vskip 5mm
\subsection{RQ7: Future work}
The last research question has to do with future work opportunities and directions. 
Our findings are summarized in Table 12 and shortly explained below:  	
\begin{description}[leftmargin=0.25cm]
	%
	\setlength\itemsep{0em}
	\item[\textbf{Extend the proposed solution}] It is a common suggestion stated by many 
	authors. They often identify and suggest several additional parts or components 
	which could be aggregated to the system to improve the performance, extend the 
	functionalities, etc. It is suggested in 14 (18.4\%) studies. 
	\item[\textbf{Perform better evaluation}] It is difficult to evaluate recommender 
	systems. The hard part is to find the most appropriate techniques or algorithms that 
	can be used as benchmark. Performing a good evaluation of the proposed system increases 
	its value and credibility. This suggestion appears in 11 (14.4\%) studies. 	
	\item[\textbf{Add context to recommendations}] The authors suggest to make more use 
	of contextual (location, time of day, etc.) data which are revealed by mobile users. 
	It appears in 8 (10.5\%) studies.	
	\item[\textbf{Consider other application domains}] Some of the studies apply their 
	contributions in a certain domain. Different authors target alternative domains or 
	propose domain independent contributions. Considering other domains was suggested 
	in 7 (9.2\%) studies. 	
	\item[\textbf{Use more data or item features}] Some authors plan to use more data 
	for training their algorithms or plan to extract and use more features of the 
	recommended items. This has been stated in 7 (9.2\%) studies.
	\item[\textbf{Experiment with more or different algorithms}] Some authors suggest 
	to combine different recommendation or data mining algorithms and see the results 
	they can obtain. Sometimes they suggest to use alternative similarity measures also. 
	This has been suggested in 6 (7.9\%) studies. 	
	\item[\textbf{Try other hybridization class}] Although it is not always possible, 
	combining the applied techniques in another way could bring better results. 
	Trying another hybridization class appeared in 5 (6.5\%) studies. 	
	\item[\textbf{Other}] Other future work suggestions include applying hybrid RSs 
	in less frequent domains or contexts, making more personalized recommendations, reducing
	the computational cost of the solution, improving other recommendation quality criteria 
	(besides accuracy) like diversity or serendipity, etc.
\end{description}
\begin{table}[ht] 
	\caption{Future work suggestions} 
	\footnotesize
	\centering       
	\begin{tabular}
		{l c}   
		\topline
		\headcol Future work & Studies  						\\ [0.5ex] 	 
		\midline 
		Extend the proposed solution & 14       				\\   		
		\rowcol Perform better evaluation & 11          		\\ 
		Other & 9       										\\ 
		\rowcol Add context to recommendations & 8     			\\
		Consider other application domains & 7           		\\ 
		\rowcol Use more data or item features & 7    			\\
		Experiment with more or different algorithms & 6           \\
		\rowcol Try other hybrid recommendation class & 5       		\\
		\bottomline
	\end{tabular}  
\end{table}
\vskip 7mm
\section{Discussion}
The main issues covered in this work are presented in the schematic model of Figure 9. 
The issues are associated with the research question they belong to. In this section 
we discuss the obtained results for each research question. 
\begin{figure}[!htbp]
	\centering
	\includegraphics[width=112mm,scale=0.5]{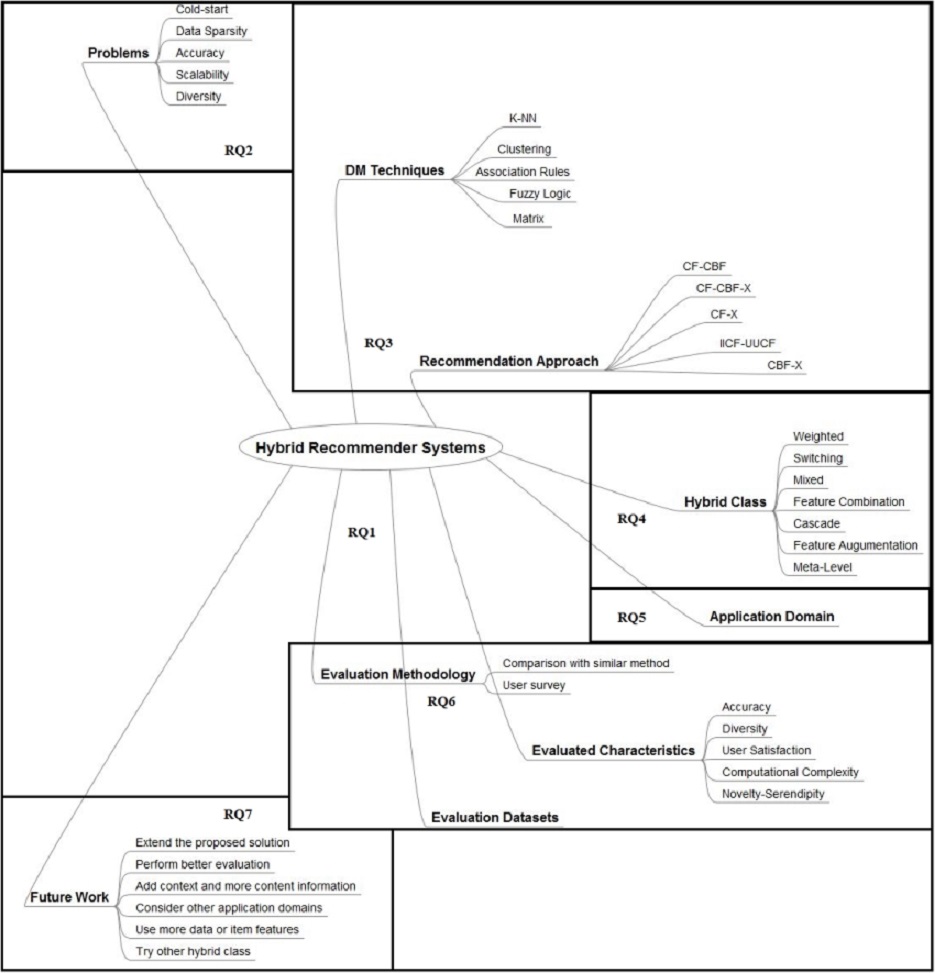}
	\caption{RQs and higher-order themes}
\end{figure}
\vskip 5mm
\subsection{Selected studies} 		
The quality evaluation results of the selected studies are presented in Figure 3 
and Figure 4. These results 
indicate that journal studies have lower spread and slightly higher quality score than 
conference studies. The authors in \cite{figueroa2015systematic}, a systematic review work 
about linked data-based recommender systems, report similar results. 
Regarding the publication year of the selected studies, we see in Figure 2 a steady 
increase in hybrid RS publications. More than 76\% of the included papers were 
published in the second half 
(from 2010 later on) of the 10 years time period. This high number of recent publications 
suggest that hybrid RSs are still a hot topic. As mentioned in introduction, similar 
increased academic interest in RSs is also reported by other surveys like 
\cite{Bobadilla:2013:RSS:2483330.2483573} or \cite{Park:2012:LRC:2181339.2181690}. 
Some factors that have boosted the publications and development of RSs are probably 
the Netflix Prize\footnote{http://www.netflixprize.com/} (2006-2009) and the boom of 
social networks.  	
\vskip 5mm
\subsection{Problems and challanges} 
Cold-start was the most acute problem that was found. 
CF RSs are the most affected by cold-start as they generate recommendations relying 
on ratings only. Hybrid RSs try to overcome the lack of ratings by combining CF or 
other recommendation techniques with association rule mining or other mathematical 
constructs which extract and use features from items. 
Data sparsity is also a very frequent problem in the field of RSs. It represents a 
recommendation quality degradation due to the insufficient number of ratings. Hybrid 
approaches try to solve it by combining several matrix manipulation techniques with 
the basic recommendation strategies. They also try to make more use of item features, 
item reviews, user demographic data or other known user characteristics. 
\par
Accuracy has been the top desired characteristic of RSs since their dawn, as it directly 
influences user satisfaction. Improving recommendation accuracy is a problem that is 
mostly addressed by using parallel (i.e. in a weighted or switching hybrid classes) 
recommendation techniques. Scalability is also an important problem which is frequently 
found in association with data sparsity (appear together in 9 studies). Lack of diversity 
is a problem that has been addressed in few studies. As explained 
in \cite{2010PNAS..107.4511Z} diversity is frequently in contradiction with accuracy. Authors 
usually attain higher diversity by tolerable relaxations in accuracy. In general we see 
that hybrid RSs try to solve the most acute problems that RSs face. In Table 13 we 
summarize some typical solutions about each problem with examples from papers discussed 
in sections 3.2 - 3.5. 
\begin{table}[ht] 
	\caption{Problems and possible solutions} 
	\footnotesize
	\centering      
	\begin{tabular}
		{p{1.3cm} p{8.4cm} p{2.4cm}} 
		\topline
		\headcol \textbf{Problems} & \textbf{Possible Solutions} & \textbf{References}  \\ 
		\midline 	\vspace*{0.1ex}	
		Cold-Start &
		\vspace*{0.1ex}	
		Use association rule mining on item or user data to find relations	 
		which can compensate the lack of ratings. Mathematical constructs for feature 
		extraction and combination of different strategies can also be used.   
		& \vspace*{0.1ex}	
		\hyperlink{P21}{[P21]},\: \hyperlink{P61}{[P61]},\: \hyperlink{P26}{[P26]}, \linebreak
		\hyperlink{58}{[P58]},\: \hyperlink{P59}{[P59]} 				\\
		\hline \\ [0.01ex] 
		Sparsity & Use the few existing ratings or certain item features to generate  
		extra pseudo ratings. Experiment with Matrix Factorization or Dimensionality Reduction. 
		& \hyperlink{P1}{[P1]},\: \hyperlink{P44}{[P44]},\: \hyperlink{P76}{[P76]}, \hyperlink{P13}{[P13]}				 \\
		\hline \\ [0.01ex]
		Accuracy & Use Fuzzy Logic or Fuzzy Clustering in association with CF. Try putting 
		together CF with CBF using Probabilistic Models, Bayesian Networks or other 
		mathematical constructs.  
		& \hyperlink{P27}{[P27]},\: \hyperlink{P34}{[P34]},\: \hyperlink{P6}{[P6]},
		\hyperlink{P40}{[P40]},\: \hyperlink{P20}{[P20]},\: \hyperlink{P51}{[P51]}		\\
		\hline \\ [0.01ex] 
		Scalability & Try to compress or reduce the datasets with Clustering or different
		measures of similarity.   
		& \hyperlink{P28}{[P28]},\: \hyperlink{P76}{[P76]},\: \hyperlink{P28}{[P28]} 	\\
		\hline \\ [0.01ex] 
		Diversity & Try modifying neighborhood creation by relaxing similarity 
		(possible loss in accuracy) or use the concept of experts for certain item tastes.  
		& \hyperlink{P36}{[P36]},\: \hyperlink{P46}{[P46]},\: \hyperlink{P12}{[P12]}		\\
		%
		\bottomline 
	\end{tabular}  
\end{table}
\vskip 5mm
\subsection{Techniques and combinations}
As shown in Table 7, K-NN is the most popular DM technique among hybrid 
RSs. This result highlights the fact that K-NN CF is one of the most successful and 
widespread RSs. Clustering techniques are also commonly used. There are different types of 
clustering algorithms with \emph{K-means} being the most popular. Clustering as a process 
is mostly involved in preliminary phases to identify similar users, similar items, similar 
item features, etc. Association rules are also used to identify frequent relations between 
users and items. Fuzzy logic and matrix manipulation methods are also incorporated in hybrid RSs.   
In most of the cases authors combine 2 recommendation strategies. In few cases event 3  
are involved. CF-CBF is the most popular combination, commonly associated 
with recurrent problems like data sparsity, cold-start and accuracy. CF-CBF-X is also  
common. Here CF and CBF are combined together and reinforced by a third technique. 
\par
In CF-X combinations, X is usually integrated in CF to improve its performance and usually 
represents fuzzy logic (reclusive methods are complementary to collaborative methods) or 
clustering. IICF-UUCF is also popular as it represents the combination of two basic version 
of CF. In conclusion, as can be inferred from Table 8, the most common recommendation 
techniques (with CF been the most popular) are combined to solve the typical problems 
which are cold-start, data sparsity and accuracy. Actually it is not a surprise that CF combines 
with almost any other recommendation technique. Other surveys report similar results. In \cite{Su:2009:SCF:1592474.1722966} the authors present a broad survey about CF techniques. 
They also conclude that most of hybrid CF recommenders use CF methods in combination with 
content-based methods (CF-CBF is also the most frequent combination we found) or other 
methods to fix problems of either recommendation technique and to improve recommendation 
performance. CBF-X addresses problems like data sparsity, accuracy and scalability. 
\par
Other combinations put together techniques like Bayesian methods, demographic filtering, 
neural networks, regression, association rules mining or genetic algorithms.  
It is important to note that in some cases hybrid RSs are not built by combining 
different recommendation techniques. In those cases they represent combinations of 
different data sources, item or user representations, etc. embedded in a single RS. 
For this reason the number of the reported combinations is smaller than the number of 
total primary studies we analyzed.
\vskip 5mm
\subsection{Hybridization classes}
Regarding the hybrid classes, weighted hybrid is the most popular. It often combines CF 
and CBF recommendations in a dynamic way (weights change over time). Feature combination 
is the second, putting together data from two or more sources. Cascade, switching, feature 
augmentation and meta-level have almost equal frequency of appearance whereas mixed hybrid 
is the least common class. There is also a last category we denoted as "Other" which 
includes 13.2\% of the studies. It was not possible for us to identify a hybridization 
class of this recommenders based on  Burke's taxonomy (which might also need to be extended). 
In some studies hybrid RSs are not combinations of two or more recommendation strategies 
in a certain way. They put together different data sources and item or user representations 
in a single strategy. In this sense, the "Other" category means "we don't know". 
\par
Various mathematical constructs are used as "gluing" methods between the different 
components of the systems based on the hybridization class. Weighted, Mixed, Switching 
and Feature Combination are order-insensitive; there is no difference between a switching
CF-CBF and a switching CBF-CF. In this sense these 4 classes are easier to concatenate 
compared to Cascade, Feature Augmentation and meta-level which are inherently ordered.   
The few mixed systems do not need the "glue" at all as their components generate 
recommendations independently from each other. Our results indicate that Weighted hybrids
usually rely on weighted linear functions with static or dynamic weights which are updated
based on the user feedback. Switching hybrids usually rely on distance/similarity measures 
such as Euclidean distance, Pearson correlation, Cosine similarity, etc. to decide which of the 
components to activate in a certain time. Feature combinations usually involve fuzzy logic 
to match the features obtained by one module with those of the other module. Feature 
augmentation, Cascade and Meta-level hybrids rely on even more complex and advanced 
mathematical frameworks such as probabilistic modeling, Bayesian networks, etc.    
\vskip 5mm
\subsection{Application domains}
A rich set of application domains was found as shown in Figure 7. Many of the studies are domain 
independent (more than a quarter). They are not limited to any particular domain and the methods 
or algorithms they present can be applied in different domains with minor or no changes at all. 
Movies are obviously the most recommended items. It is somehow because of the large 
amount of public and freely accessible user feedback about movie preferences (i.e. 
many public movie datasets on the web\footnote{\url{https://gist.github.com/entaroadun/1653794}}) 
which are highly helpful. There is also a rich set of algorithms and solutions (Netflix \$1M challenge 
was a big motivation to improve movie recommenders). This allows researchers to train and test their 
recommendation algorithms easily. Education or e-learning is another domain in which hybrid RSs are 
gaining popularity. The amount of educational material on the web has been increasing dramatically
in the last years and MOOCs (Massive Open Online Course) are becoming very popular. Other somehow 
popular domains are music and web services. More detailed information about the application domains 
of recommender systems can be found at \cite{D174} where the authors illustrate each application 
domain category with real RS applications found in the web.   
\vskip 5mm
\subsection{Evaluation}
Evaluation of Recommender Systems is an essential phase which helps in choosing the right 
algorithm in a certain context and for a certain problem. However, as explained in 
\cite{Herlocker:2004:ECF:963770.963772}, evaluating recommender systems is not an easy 
task. Certain algorithms may perform better or worse in different datasets and it is not 
easy to decide what metrics to combine when performing comparative evaluations. 
With the three research questions about evaluation, we addressed different aspects of this 
delicate process. Based on our results most of the studies evaluate hybrid RSs by comparing 
them with similar methods. The experiments which are usually offline utilize accuracy or error 
metrics like MAE or RMSE and information retrieval metrics like precision, recall and F1.
Similar results are reported in \cite{Beel2015r} where offline evaluations that typically 
measure accuracy are dominant. User surveys are less popular, using subjective quality 
assessments and occasionally precision or recall. These kind of experiments are mostly 
online (i.e. users interacting with the system and answering questions) and offer more
direct and credible evaluation conclusions. From the results, we see that researchers find 
it easier to compare their system with other systems using public data rather than to perform 
massive user surveys for a more subjective and qualitative evaluation. 
\par
Regarding RS characteristics, accuracy results to be the most commonly 
evaluated characteristic of the hybrid RSs. This is partly because it is easy to 
represent and compute it by means of various measures that exist. The most frequent 
metrics used to evaluate accuracy are Precision, Recall and MAE.  
User satisfaction (subjective recommendation quality) comes second. It is  
evaluated by means of user surveys. There is a lot of discussion in the literature about 
recommendation diversity. In \cite{Ziegler:2005:IRL:1060745.1060754} the authors conclude 
that the user's overall liking of recommendations goes beyond accuracy and involves other 
factors like diversity. On the other hand, in \cite{2010PNAS..107.4511Z} the authors 
agree that increasing diversity in recommendations comes with a cost in accuracy. 
Our results show that diversity is still less frequently evaluated. Actually most of the 
studies that try to provide diversity do it by conceding accuracy. In 
\cite{Ge:2010:BAE:1864708.1864761} the authors explore the use of serendipity and coverage as 
both characteristics and quality measures of RSs. They suggest that serendipity and coverage 
are designed to account for the quality and usefulness of the recommendations better than 
accuracy does. In our results serendipity is rarely evaluated.
\par
It is important to note that the difference between recommendation characteristics and 
evaluation metrics is sometimes subtle. This is the case for coverage. Is coverage a 
recommendation characteristic, a recommendation metric or both? In some works like 
\cite{Herlocker:2004:ECF:963770.963772} and \cite{Ge:2010:BAE:1864708.1864761} coverage is 
considered as both a characteristic and metric. As a characteristic it reflects the usefulness 
of the system. The higher the coverage (more items predicted and recommended) the more 
useful the recommender system for the users. In other works like \cite{5680904} it is only 
considered as a metric with which the authors evaluate diversity, another recommendation 
characteristic. In the studies we considered for this review coverage is both considered as
a metric for estimating the diversity and as a recommendation characteristic of the systems.
Few studies we analyzed evaluate the computational complexity of the systems they propose by 
measuring the execution time. Besides the new trends, the results indicate that accuracy is 
still the most frequently evaluated characteristic. 
\par
We also considered the public datasets 
used to perform the evaluation. With the exponential growth of the web content there are 
more and more public data and datasets which can be used 
to train and test new algorithms. These datasets usually come from highly visited web portals 
or services and represent user preferences about things like movies, music, news, books, etc.
In \cite{7325106} we present the characteristics of some of the most popular public datasets 
and the types of RSs they can be used for. It is convenient to exploit them for evaluating 
novel algorithms or recommendation techniques in offline experiments. The evaluation process 
steps are clearly explained in \cite{export:115396}. The result of this review indicate that movie 
datasets led by Movielens are very popular being used in more than 72\% of the studies. This 
is somehow related with the fact that movie domain is also highly preferred. Many authors 
chose to experiment in the domain of movies to easily evaluate their prototypes. Music, web 
services, tourism, images datasets, etc. make up the rest of the datasets the studies use.  
\vskip 5mm
\subsection{Future work}
With RQ7 we tried to uncover the most important future work directions in hybrid recommender 
systems.
Extending or improving the proposed solution is the most common future work the authors
intend to undertake. Extension of the proposed solutions comes in diverse forms like 
\textit{(i)} extend by applying more algorithms, \textit{(ii)} extend the personalization 
level by adapting more to the user context and profile, \textit{(iii)} extend by using 
more datasets or item features, etc. 
Performing a comprehensive evaluation is something 
in which many studies fail. This is why some authors present it as a future work. It usually 
happens in the cases when the authors implement their algorithm or method in a prototype. 
In these cases comparison with similar methods using accuracy metrics does not provide 
clear insights about recommendation or system quality. Reinforcing with subjective user 
feedback may be the best way to optimize evaluation of the system, making it more user
oriented. 
\par
A highly desired characteristic from RSs is adapting to the user interest shifting or 
evolving over time, especially as a results of rapid context changes. As a result,
different authors suggest to add context to their systems or to analyze different 
criteria of items or users as ways to improve the recommendation quality. 
Context-Aware Recommender Systems (CARS) and Multi-Criteria Recommender Systems 
(MCRS) are relatively new approaches which are gaining popularity 
in the field of RSs \cite{Adomavicius2011}. They are promoted by the increased use of 
mobile devices which reveal user details (i.e. the location) that can be used as 
important contextual inputs. Combining context and multiple criteria with other hybrid 
recommendation techniques could be a good direction in which to experiment.
\par
Considering other application domains in which hybrid RSs could be applied is also stated
by some authors. Many of the works were domain independent and can be easily adapted to 
different recommendation domains.
One step further could be to have hybrid RSs recommend items from different (changing) 
domains and implement the so called cross domain recommender systems. 
Having found the best movie for the weekend, the user may also want to find the 
corresponding soundtrack or the book in which the movie may be based on. Cross-domain RSs 
are an emerging research topic \cite{Cremonesi:2011:CRS:2117693.2119484,fernandez2012cross}. 
Different recommendation strategies like CF and CBF could be specialized in different 
domains of interest and then joined together in a weighted, switching, mixed or other 
hybrid cross-domain RS which would recommend different items to its users.
\par
Combining more data from different sources or with various item features was a way to 
create hybrid RSs. Using more data is a common trend not only in recommender systems but 
in similar disciplines as well. However, having and using big volumes of data requires 
scaling in computations. One way to achieve this high scalability is by parallelizing the 
algorithms following \emph{MapReduce} model which could be a future direction as suggested in 
\cite{xyz}. 
Experimenting with other hybrid recommendation classes is also possible in many cases. 
The results indicate that some hybrid classes are rarely explored (i.e. mixed hybrid appears 
in 3 studies only). It could be a good idea to experiment building CF-CBF, CF-CBF, CF-KBF or 
other types of mixed hybrids and observe what characteristics this systems could provide.  
Other future work suggestions include increasing personalization and reducing the computational 
cost of the system.  
\vskip 7mm
\section{Conclusions} 
In this review work we analyzed 76 primary studies from journals and conference proceedings
which address hybrid RSs. We tried to identify the most acute problems they solve to 
provide better recommendations. We also analyzed the data mining and machine learning 
techniques they use, the recommendation strategies they combine, hybridization 
classes they belong to, application domains and dataset, evaluation process, and possible 
future work directions.
\par
With regard to the research problems cold-start, data sparsity and accuracy are 
the most recurrent problems for which hybrid approaches are explored. The authors 
typically use association rules mining in combination with traditional recommendation 
strategies to find user-item relations and compensate the lack of ratings in cold-start 
situations. We also found that matrix factorization techniques help to compress the 
existing sparse ratings and attain acceptable accuracy. It was also typical to find 
studies in which collaborative filtering was combined with other techniques such as 
fuzzy logic attempting to alleviate cold-start or data sparsity and at the same 
time provide good recommendation accuracy.         
\par
We also presented a classification of the included studies based on the different DM/ML
techniques they utilize to build the systems and their recommendation technique combinations.  
K-NN classifier which is commonly used to construct the neighborhood in collaborative RSs, 
was the most popular among the data mining technique. On the other hand, CF was the
most commonly used recommendation strategy, frequently combined with each of the other 
strategies attempting to solve any kind of problem. 
\par
We identified and classified the different hybridization approaches relying in the taxonomy 
proposed by Burke and found that the weighted hybrid is the most recurrent, obviously 
because of the simplicity and dynamicity it offers. Other hybridization classes such as
meta level or feature augmentation are rare as they need complicated mathematical 
constructs to aggregate the results of the different recommenders they combine. 
\par	
Concerning evaluation, accuracy is still considered the most important characteristic. 
The authors predominantly use comparisons with similar methods and involve error or prediction 
metrics in the evaluation process. This evaluation methodology is "hermetic" and often not
credible. User satisfaction is commonly evaluated with subjective data feedback from  
surveys which are user oriented, more credible and thus highly suggested. Additionally, 
computational complexity was found in few cases. We also investigated what public datasets
are typically used to perform evaluation of the hybrid systems. Based on our findings
movie datasets led by Movielens are the most popular, facilitating the evaluation 
process. Moreover movie domain was the most preferred for prototyping, among the numerous 
that were identified. 
\par
More than three-quarters of our included studies were published in the last five years.  
This high and growing number of recent publications in the field lets us believe that hybrid 
RSs are a hot and interesting topic. Our findings indicate that future works could be focused 
in context awareness of recommendations and models with which to formalize and aggregate 
severals contextual factors inside a hybrid recommender. Such RSs could be able to respond 
to quick shifts of user interest with high accuracy. 
\par
We also found that there are many combinations of recommendation techniques or 
hybridization classes which are not explored. Thus they represent a good basis for 
future experimentations in the field. Using more data was another possible work 
direction we found. In the epoch of big data, processing more or larger 
dataset (as even more become available) with hybrid parallel algorithms could 
be a good way to alleviate the problem of scalability and also provide better 
recommendation quality. Other future work direction could be using hybrid RSs to 
build cross domain recommenders or improve the computation complexity of the 
existing techniques.  
\vskip 7mm
\section*{Acknowledgments}
This work was supported by a fellowship from TIM\footnote{https://www.tim.it/}.
%
\vskip 7mm
\section*{Appendix A. Selected Papers}
%
\newcolumntype{L}[1]{>{\raggedright\let\newline\\\arraybackslash\hspace{0pt}}p{#1}}
\begingroup
\fontsize{6pt}{7pt}\selectfont
\begin{longtable}{l L{2cm} l L{3.8cm} L{0.7cm} L{3.8cm}}
	\caption{Selected papers} \\
	\topline
	\headcol P & Authors & Year & Title & Source & Publication details \\
	\midline
	\endfirsthead
	\multicolumn{6}{l}
	{\tablename\ \thetable\ -- \textit{Continued from previous page}} \\
	\hline
	\headcol P & Authors & Year & Title & Source & Publication details \\
	\midline
	\endhead
	\hline \multicolumn{6}{l}{\textit{Continued on next page}} \\
	\endfoot
	\hline
	\endlastfoot
	\hypertarget{P1}{P1} & Wang, J.; \newline De Vries, P. A.; Reinders, J. T. M.; & 2006 & Unifying User-based and Item-based Collaborative Filtering Approaches by Similarity Fusion  & ACM & 29th Annual International ACM SIGIR Conference on Research \& Development on Information Retrieval, Seattle 2006 \\ [1ex]
	
	\rowcol \hypertarget{P2}{P2} & Gunawardana, A.; Meek, C.; & 2008 & Tied Boltzmann Machines for Cold Start Recommendations & ACM 
	& 2nd ACM Conference on Recommender Systems, Lousanne, Switzerland, 23rd-25th October 2008 \\ [1ex]
	
	P3 & Gunawardana, A.; Meek, C.; & 2009 & A Unified Approach to Building Hybrid Recommender Systems & ACM 
	& 3rd ACM Conference on Recommender Systems, New York, October 23-25, 2009 \\ [1ex]
	
	\rowcol P4 & Park, S. T.; Chu, W.; & 2009 & Pairwise Preference Regression for Cold-start Recommendation & ACM & 3rd ACM Conference on Recommender Systems, 
	New York, October 23-25, 2009  \\ [1ex]
	
	\hypertarget{P5}{P5} & Ghazanfar, M. A.; Prugel-Bennett, A.; & 2010 & An Improved Switching Hybrid Recommender System Using Naive Bayes Classifier and 
	Collaborative Filtering & ACM & Proceedings of the International MultiConference of Engineers and Computer Scientists 2010, Vol I, Hong Kong, 
	March 17-19, 2010 \\ [1ex]
	
	\rowcol \hypertarget{P6}{P6} & Zhuhadar, L.; Nasraoui, O.; & 2010 & An Improved Switching Hybrid Recommender System Using Naive Bayes Classifier and 
	Collaborative Filtering & ACM & Proceedings of the International MultiConference of Engineers and Computer Scientists 2010, Vol I, Hong Kong, 
	March 17-19, 2010 \\[1ex]
	
	\hypertarget{P7}{P7} & Hwang, C. S.; & 2010 & Genetic Algorithms for Feature Weighting in Multi-criteria Recommender Systems & ACM & Journal of Convergence 
	Information Technology, Vol. 5, N. 8, October 2010 \\[1ex]
	
	\rowcol \hypertarget{P8}{P8} & Liu, L.; Mehandjiev, N.; Xu, D. L.; & 2011 & Multi-Criteria Service Recommendation Based on User Criteria Preferences
	& ACM & 5th ACM Conference on Recommender Systems, Chicago, Oct 23rd-27th 2011 \\[1ex]
	
	P9 & Bostandjiev, S.; O’Donovan, J.; Höllerer, T.; & 2012 & TasteWeights: A Visual Interactive Hybrid Recommender System & ACM & 6th ACM Conference on Recommender Systems, Dublin, Sep. 9th-13th, 2012 \\[1ex]
	
	\rowcol \hypertarget{P10}{P10} & Stanescu, A.; Nagar, S.; Caragea, D.; & 2013 & A Hybrid Recommender System: User Profiling from Keywords and Ratings
	& ACM & A Hybrid Recommender System: User Profiling from Keywords and Ratings \\ [1ex]
	
	\hypertarget{P11}{P11} & Hornung, T.; Ziegler, C. N.; Franz, S.; & 2013 & Evaluating Hybrid Music Recommender Systems & ACM & 2013 IEEE/WIC/ACM International 
	Conferences on Web Intelligence (WI) and Intelligent Agent Technology (IAT) \\ [1ex]
	
	\rowcol \hypertarget{P12}{P12} & Said, A.; Fields, B.; Jain, B. J.; & 2013 & User-Centric Evaluation of a K-Furthest Neighbor Collaborative Filtering 
	Recommender Algorithm & ACM & The 16th ACM Conference on Computer Supported Cooperative Work and Social Computing, Texas, Feb. 2013 \\ [1ex]
	
	\hypertarget{P13}{P13} & Hu, L.; Cao, J.; Xu, G.; Cao, L.; Gu, Z.; Zhu, C.; & 2013 & Personalized Recommendation via Cross-Domain Triadic 
	Factorization & Scopus & 22nd ACM International WWW Conference, May 2013, Brasil \\ [1ex]
	
	\rowcol \hypertarget{P14}{P14} & Christensen, I.; Schiaffino, S.; & 2014 & A Hybrid Approach for Group Profiling in Recommender Systems 
	& ACM & Journal of Universal Computer Science, vol. 20, no. 4, 2014 \\ [1ex]
	
	P15 & Garden, M.; Dudek, G.; & 2005 & Semantic feedback for hybrid recommendations in Recommendz & IEEE & IEEE 2005 International 
	Conference on e-Technology, e-Commerce and e-Service \\ [1ex]
	
	\rowcol P16 & Bezerra, B. L. D.; Carvalho, F. T.; Filho, V. M.; & 2006 & C2 :: A Collaborative Recommendation System Based on Modal 
	Symbolic User Profile & IEEE & Proceedings of the 2006 IEEE/WIC/ACM International Conference on Web Intelligence \\ [1ex]
	
	P17 & Ren, L.; He, L.; Gu, J.; Xia, W.; Wu, F.; & 2008 & A Hybrid Recommender Approach Based on Widrow-Hoff Learning & IEEE & IEEE 2008 Second International Conference on Future Generation Communication and Networking \\ [1ex]
	
	\rowcol P18 & Godoy, D.; Amandi, A.; & 2008 & Hybrid Content and Tag-based Profiles for Recommendation in Collaborative Tagging 
	Systems & IEEE & IEEE 2008 Latin American Web Conference \\ [1ex]
	
	P19 & Aimeur, E.; Brassard, G.; Fernandez, J. M.; Onana, F. S. M.; Rakowski, Z.; & 2008 & Experimental Demonstration of 
	a Hybrid Privacy-Preserving Recommender System & IEEE & The Third International Conference on Availability, Reliability and Security, 
	IEEE 2008 \\ [1ex]
	
	\rowcol \hypertarget{P20}{P20} & Yoshii, K.; Goto, M.; Komatani, K.; Ogata, T.; Okuno, H. G.; & 2008 & An Efficient Hybrid Music Recommender System Using 
	an Incrementally Trainable Probabilistic Generative Model & IEEE & IEEE TRANSACTIONS ON AUDIO, SPEECH, AND LANGUAGE PROCESSING, VOL. 
	16, NO. 2, FEBRUARY 2008 \\ [1ex]
	
	\hypertarget{P21}{P21} & Maneeroj, S.; Takasu, A.; & 2009 & Hybrid Recommender System Using Latent Features & IEEE & IEEE 2009 International Conference 
	on Advanced Information Networking and Applications \\ [1ex]
	
	\rowcol P22 & Meller, T.; Wang, E.; Lin, F.; Yang, C.; & 2009 & New Classification Algorithms for Developing Online Program
	Recommendation Systems & IEEE & IEEE 2009 International Conference on Mobile, Hybrid, and On-line Learning \\ [1ex]
	
	\hypertarget{P23}{P23} & Shambour, Q.; Lu, J.; & 2010 & A Framework of Hybrid Recommendation System for Government-to-Business Personalized 
	e-Services & IEEE & IEEE 2010 Seventh International Conference on Information Technology \\ [1ex]
	
	\rowcol \hypertarget{P24}{P24} & Deng, Y.; Wu, Z.; Tang, C.; Si, H.; Xiong, H.; Chen, Z.; & 2010 & A Hybrid Movie Recommender Based on Ontology and 
	Neural Networks & IEEE & A Hybrid Movie Recommender Based on Ontology and Neural Networks \\ [1ex]
	
	\hypertarget{P25}{P25} & Yang, S. Y.; Hsu, C. L.; & 2010 & A New Ontology-Supported and Hybrid  Recommending Information System for Scholars & Scopus & 13th International Conference on Network-Based Information Systems \\ [1ex]
	
	\rowcol \hypertarget{P26}{P26} & Basiri, J.; Shakery, A.; Moshiri, B.; Hayat, M.; & 2010 & Alleviating the Cold-Start Problem of Recommender Systems 
	Using a New Hybrid Approach & IEEE & IEEE 2010 5th International Symposium on Telecommunications (IST'2010) \\ [1ex]
	
	\hypertarget{P27}{P27} & Valdez, M. G.; Alanis, A.; Parra, B.; & 2010 & Fuzzy Inference for Learning Object Recommendation & IEEE & IEEE 2010 
	International Conference on Fuzzy Systems \\ [1ex]
	
	\rowcol \hypertarget{P28}{P28} & Choi, S. H.; Jeong, Y. S.; Jeong, M. K.; & 2010 & A Hybrid Recommendation Method with Reduced Data for Large-Scale 
	Application & IEEE & IEEE Transactions on systems, man and cybernetics - Part C: Applicatios and Reviews, VOL. 40, NO. 5, 
	September 2010 \\ [1ex]
	
	\hypertarget{P29}{P29} & Ghazanfar, M. A.;  Prugel-Bennett, A.; & 2010 & Building Switching Hybrid Recommender System Using Machine Learning 
	Classifiers and Collaborative Filtering & IEEE & IEEE IAENG International Journal of Computer Science, 37:3, IJCS\_37\_3\_09 \\ [1ex]
	
	\rowcol P30 & Castro-Herrera, C.; & 2010 & A Hybrid Recommender System for Finding Relevant Users in Open Source Forums & Scopus & IEEE 3rd International Conference on Managing Requirements Knowledge, 
	Sept. 2010 \\ [1ex]
	
	P31 & Tath, I.; Biturk, A.; & 2011 & A Tag-based Hybrid Music Recommendation System Using Semantic Relations and Multi-domain 
	Information & IEEE & 11th IEEE International Conference on Data Mining Workshops, Dec. 2011 \\ [1ex]
	
	\rowcol P32 & Kohi, A.; Ebrahimi, S. J.; Jalali, M.; & 2011 & Improving the Accuracy and Efficiency of Tag Recommendation System by
	Applying Hybrid Methods & IEEE & IEEE 1st International eConference on Computer and Knowledge Engineering (ICCKE), October 13-14, 
	2011 \\ [1ex]
	
	P33 & Kohi, A.; Ebrahimi, S. J.; Jalali, M.; & 2011 & Improving the Accuracy and Efficiency of Tag Recommendation System by
	Applying Hybrid Methods & IEEE & IEEE 1st International eConference on Computer and Knowledge Engineering (ICCKE), October 13-14, 
	2011 \\ [1ex]
	
	\rowcol \hypertarget{P34}{P34} & Fenza, G.; Fischetti, E.; Furno, D.; Loia, V.; & 2011 & A hybrid context aware system for tourist guidance based on collaborative filtering & Scopus & 2011 IEEE International Conference on Fuzzy Systems, 
	June 27-30, 2011, Taipei, Taiwan \\ [1ex]
	
	P35 & Shambour, Q.; Lu, J.; & 2011 & A Hybrid Multi-Criteria Semantic-enhanced Collaborative Filtering Approach for 
	Personalized Recommendations & IEEE & 2011 IEEE/WIC/ACM International Conferences on Web Intelligence and Intelligent Agent 
	Technology \\ [1ex]
	
	\rowcol \hypertarget{P36}{P36} & Li, X.; Murata, T.; & 2012 & Multidimensional Clustering Based Collaborative Filtering Approach for Diversified 
	Recommendation & IEEE & The 7th International Conference on Computer Science \& Education July 14-17, 2012. Melbourne, 
	Australia \\ [1ex]
	
	\hypertarget{P37}{P37} & Shahriyary, S.; Aghabab, M. P.; & 2013 & Recommender systems on web service selection problems using a new hybrid 
	approach & IEEE & IEEE 4th International Conference on Computer and Knowledge Engineering, 2014 \\ [1ex]
	
	\rowcol \hypertarget{P38}{P38} & Yu, C. C.; Yamaguchi, T.; Takama, Y.; & 2013 & A Hybrid Recommender System based Non-common Items in Social Media & IEEE & IEEE International Joint Conference on Awareness Science and Technology and Ubi-Media Computing, 2013 \\ [1ex]
	
	P39 & Buncle, J.; Anane, R.; Nakayama, M.; & 2013 & A Recommendation Cascade for e-learning & IEEE & 2013 IEEE 27th International 
	Conference on Advanced Information Networking and Applications \\ [1ex]
	
	\rowcol \hypertarget{P40}{P40} & Bedi, P.; Vashisth, P.; Khurana, P.; & 2013 & Modeling User Preferences in a Hybrid Recommender System using Type-2 Fuzzy Sets & Scopus & IEEE International Conference on Fuzzy Systems, July 2013 \\ [1ex]
	
	\hypertarget{P41}{P41} & Andrade, M. T.; Almeida, F.; & 2013 & Novel Hybrid Approach to Content Recommendation based on Predicted 
	Profiles & IEEE & 2013 IEEE 10th International Conference on Ubiquitous Intelligence \& Computing \\ [1ex]
	
	\rowcol P42 & Yao, L.; Sheng, Q. Z.; Segev, A.; Yu, J.; & 2013 & Recommending Web Services via Combining Collaborative Filtering 
	with Content-based Features & IEEE & 2013 IEEE 20th International Conference on Web Services \\ [1ex]
	
	P43 & Luo, Y.; Xu, B.; Cai, H.; Bu, F.; & 2014 & A Hybrid User Profile Model for Personalized Recommender System with Linked 
	Open Data & IEEE & IEEE 2014 Second International Conference on Enterprise Systems \\ [1ex]
	
	\rowcol \hypertarget{P44}{P44} & Sharif, M. A.; Raghavan, V. V.; & 2014 & A Clustering Based Scalable Hybrid Approach for Web Page 
	& IEEE & 2014 IEEE International Conference on Big Data \\ [1ex]
	
	P45 & Xu, S.; Watada, J.; & 2014 & A Method for Hybrid Personalized Recommender based on Clustering of Fuzzy User Profiles & IEEE & IEEE International Conference on Fuzzy Systems (FUZZ-IEEE) July 6-11, 2014, Beijing, China \\ [1ex]
	
	\rowcol \hypertarget{P46}{P46} & Lee, K.; Lee, K.; & 2014 & Using Dynamically Promoted Experts for Music Recommendation & IEEE & IEEE Transactions on 
	Multimedia, VOL. 16, NO. 5, August 2014 \\ [1ex]
	
	\hypertarget{P47}{P47} & Chughtai, M. W.; Selamat, A.; Ghani, I.; Jung, J. J.; & 2014 & E-Learning Recommender Systems Based on Goal-Based 
	Hybrid Filtering & IEEE & International Journal of Distributed Sensor Networks Volume 2014pages \\ [1ex]
	
	\rowcol P48 & Li, Y.; Lu, L.; Xufeng, L. & 2005 & A hybrid collaborative filtering method for multiple-interests and 
	multiple-content recommendation in E-Commerce & Science Direct & Expert Systems with Applications 28 (2005) 67–77 \\ [1ex]
	
	\hypertarget{P49}{P49} & Kunaver, M.; Pozrl, T.; Pogacnik, M.; Tasic, J.; & 2007 & Optimisation of combined collaborative recommender 
	systems & Science Direct & International Journal of Electronics and Communications (AEU), 2007, 433-443 \\ [1ex]
	
	\rowcol P50 & Albadvi, A.; Shahbazi, M.; & 2009 & A hybrid recommendation technique based on product category attributes & Scopus & Expert Systems with Applications 36 (2009) 11480–11488 \\ [1ex]
	
	\hypertarget{P51}{P51} & Capos, L. M.; Fernandez-Luna, J. M.; Huete, J. F.; Rueda-Morales, M. A.; & 2010 & Combining content-based and 
	collaborative recommendations: A hybrid approach based on Bayesian networks & Science Direct & International Journal of Approximate 
	Reasoning 51 (2010) 785–799 \\ [1ex]
	
	\rowcol \hypertarget{P52}{P52} & Barragans-Martínez, A. B.; Costa-Montenegro, E.; Burguillo, J. C.; Rey-Lopez, M.; Mikic-Fonte, F. A.; Peleteiro, A.; 
	& 2010 & A hybrid content-based and item-based collaborative filtering approach to recommend TV programs enhanced with 
	singular value decomposition & Science Direct & International Journal of Information Sciences 180 (2010) 4290–4311 \\ [1ex]
	
	\hypertarget{P53}{P53} & Wen, H.; Fang, L.; Guan, L.; & 2012 & A hybrid approach for personalized recommendation of news on the Web
	& Science Direct & International Journal of Expert Systems with Applications 39 (2012) 5806–5814 \\ [1ex]
	
	\rowcol P54 & Porcel, C.; Tejeda-Lorente, A.; Martinez, M. A.; Herrera-Viedma, E.; & 2012 & A hybrid recommender system for 
	the selective dissemination of research resources in a Technology Transfer Office & Science Direct & International Journal of Information 
	Sciences 184 (2012) 1–19 \\ [1ex]
	
	\hypertarget{P55}{P55} & Noguera, J. M.; Barranco, M. J.; Segura, R. J.; Martinez, L.; & 2012 & A mobile 3D-GIS hybrid recommender system for tourism & Science Direct & International Journal of Information Sciences 215 (2012) 37–52 \\ [1ex]
	
	\rowcol P56 & Salehi, M.; Pourzaferani, M.; Razavi, S. A.; & 2013 & Hybrid attribute-based recommender system for learning 
	material using genetic algorithm and a multidimensional information model & Science Direct & Egyptian Informatics Journal (2013) 
	14, 67–78 \\ [1ex]
	
	P57 & Zang, Z.; Lin, H.; Liu, K.; Wu, D.; Zhang, G.; Lu, J.; & 2013 & A hybrid fuzzy-based personalized recommender 
	system for telecom products/services & Science Direct & International Journal of Information Sciences 235 (2013) 117–129 \\ [1ex]
	
	\rowcol \hypertarget{P58}{P58} & Kardan, A. A.; Ebrahimi, M.; & 2013 & A novel approach to hybrid recommendation systems based on association rules 
	mining for content recommendation in asynchronous discussion groups & Science Direct & International Journal of Information Sciences 219 
	(2013) 93–110 \\ [1ex]
	
	\hypertarget{P59}{P59} & Lucas, J. P.; Luz, N.; Moreno, M. N.; Anacleto, R.; Figueiredo, A. A.; Martins, C.; & 2013 & A hybrid recommendation 
	approach for a tourism system & Science Direct & International Journal of Expert Systems with Applications 40 (2013) 3532–3550 \\ [1ex]
	
	\rowcol \hypertarget{P60}{P60} & Son, L. H.; & 2014 & HU-FCF: A hybrid user-based fuzzy collaborative filtering method in Recommender Systems & Science Direct & International Journal of Expert Systems with Applications 41 (2014) 6861–6870 \\ [1ex]
	
	\hypertarget{P61}{P61} & Son, L. H.; & 2014 & HU-FCF++: A novel hybrid method for the new user cold-start problem in recommender systems & Scopus & Engineering Applications of Artificial Intelligence 41(2015)207–222 \\ [1ex]
	
	\rowcol P62 & Lekakos, G.; Caravelas, P.; & 2006 & A hybrid approach for movie recommendation & Springer & Multimed Tools Appl (2008) 36:55–70 DOI 10.1007/s11042-006-0082-7, Springer \\ [1ex]
	
	\hypertarget{P63}{P63} & Lekakos, G.; Giaglis, G. M.; & 2007 & A hybrid approach for improving predictive accuracy of collaborative 
	filtering algorithms & Springer & User Model User-Adap Inter (2007) 17:5–40 DOI 10.1007/s11257-006-9019-0, Springer \\ [1ex]
	
	\rowcol P64 & Degemmis, M.; Lops, P.; Semeraro, G.; & 2007 & A content-collaborative recommender that exploits WordNet-based 
	user profiles for neighborhood formation & Springer & User Model User-Adap Inter (2007) 17:217–255, DOI 10.1007/s11257-006-9023-4, 
	Springer \\ [1ex]
	
	P65 & Cho, J.; Kang, E.; & 2010 & Personalized Curriculum Recommender System Based on Hybrid Filtering & Springer & ICWL 2010, 
	LNCS 6483, pp. 62–71, 2010, Springer \\ [1ex]
	
	\rowcol P66 & Aksel, F.; Biturk, A.; & 2010 & Enhancing Accuracy of Hybrid Recommender Systems through Adapting the Domain Trends & Scopus & Workshop on the Practical Use of Recommender Systems, Algorithms and Technologies held in conjunction with RecSys 2010. Sept. 30, 2010, Barcelona \\ [1ex]
	
	\hypertarget{P67}{P67} & Lampropoulos, A. S.; Lampropoulos, P. S.; Tsihrintzis, G. A.; & 2011 & A Cascade-Hybrid Music Recommender 
	System for mobile services based on musical genre classification and personality diagnosis & Springer & Multimed Tools Appl 
	(2012) 59:241–258 DOI 10.1007/s11042-011-0742-0, Springer \\ [1ex]
	
	\rowcol \hypertarget{P68}{68} & Chen, W.; Niu, Z.; Zhao, X.; Li, Y.; & 2012 & A hybrid recommendation algorithm adapted in e-learning 
	environments & Springer & World Wide Web (2014) 17:271–284 DOI 10.1007/s11280-012-0187-z \\ [1ex]
	
	P69 & Sanchez, F.; Barrileo, M.; Uribe, S.; Alvarez, F.; Tena, A.; Mendez, J. M.; & 2012 & Social and Content Hybrid 
	Image Recommender System for Mobile Social Networks & Springer & Mobile Netw Appl (2012) 17:782–795 DOI 10.1007/s11036-012-0399-6, 
	Springer \\ [1ex]
	
	\rowcol \hypertarget{P70}{P70} & Zheng, X.; Ding, W.; Xu, J.; Chen, D.; & 2013 & Personalized recommendation based on review topics & Scopus & SOCA (2014) 8:15–31 DOI 10.1007/s11761-013-0140-8 \\ [1ex]
	
	P71 & Cao, J.; Wu, Z.; Wang, Y.; Zhuang, Y.; & 2013 & Hybrid Collaborative Filtering algorithm for bidirectionalWeb service recommendation & Springer & Knowl Inf Syst (2013) 36:607–627 DOI 10.1007/s10115-012-0562-1 \\ [1ex]
	
	\rowcol P72 & Burke, R.; Vahedian, F.; Mobasher, B.; & 2014 & Hybrid Recommendation in Heterogeneous Networks 
	& Springer & UMAP 2014, LNCS 8538, pp. 49–60, 2014, Springer \\ [1ex]
	
	P73 & Nikulin, V.; & 2014 & Hybrid Recommender System for Prediction of the Yelp Users Preferences
	& Springer & ICDM 2014, LNAI 8557, pp. 85–99, 2014, Springer \\ [1ex]
	
	\rowcol P74 & Sarne, G. M. L.; & 2014 & A novel hybrid approach improving effectiveness of recommender systems
	& Springer & J Intell Inf Syst DOI 10.1007/s10844-014-0338-z \\ [1ex]
	
	\hypertarget{P75}{P75} & Zhao, X.; Niu, Z.; Chen, W.; Shi, C.; Niu, K.; Liu, D.; & 2014 & A hybrid approach of topic model and matrix 
	factorization based on two-step recommendation framework & Springer & J Intell Inf Syst DOI 10.1007/s10844-014-0334-3, 
	Springer \\ [1ex]
	
	\rowcol \hypertarget{P76}{P76} & Nilashi, M.; Ibrahim, O. B.; Ithnin, N.; Zakaria, R.; & 2014 & A multi-criteria recommendation system using 
	dimensionality reduction and Neuro-Fuzzy techniques & Springer & Soft Comput DOI 10.1007/s00500-014-1475-6, Springer-Verlag 
	Berlin Heidelberg 2014 \\ [1ex]
	\bottomline
\end{longtable}
\endgroup
%
\vskip 3mm
\section*{References}

\bibliographystyle{elsarticle-num} 
\bibliography{HRS-SLR}





\end{document}